\documentclass[letterpaper, twocolumn]{article}
\usepackage{usenix}

\usepackage[table,dvipsnames]{xcolor}

\usepackage{graphicx}
\usepackage{tikz}
\usepackage{pgfplots}
\pgfplotsset{width=8cm,compat=1.9}

\usepackage{amsmath}
\usepackage{amssymb}

\usepackage{float}
\usepackage{subcaption}
\usepackage{caption}
\usepackage{multirow}
\usepackage{multicol}
\usepackage{booktabs}
\usepackage{makecell}
\usepackage{tabularx}
\usepackage{longtable}
\usepackage{threeparttable}
\usepackage{wrapfig}
\usepackage{ulem}
\usepackage{diagbox}
\usepackage{adjustbox}
\usepackage{arydshln}
\usepackage{tcolorbox}

\usepackage{csquotes}
\usepackage{soul}
\usepackage{enumitem}
\usepackage{balance}
\usepackage{url}
\usepackage{breqn}
\usepackage{bbding}
\usepackage{pifont}

\usepackage{hyperref}
\usepackage[ruled,vlined]{algorithm2e}

\usepackage[available,functional]{usenixbadges}
\renewcommand\footnotemark{}
\newcommand{\mkletter}[0]{{\color{LimeGreen}{\Envelope}}}
\newcommand{\circled}[1]{%
  \begin{tikzpicture}[baseline=(char.base)]
    \node[shape=circle, draw, inner sep=1pt] (char) {#1};
  \end{tikzpicture}%
}

\DeclareMathOperator*{\argmin}{arg\,min}

\begin{document}

\title{\Large \bf SafeSpeech: Robust and Universal Voice Protection Against \\Malicious Speech Synthesis}

\author{
{\rm Zhisheng Zhang}$^1$, 
{\rm Derui Wang}$^2$ \mkletter \rm ,
{\rm Qianyi Yang}$^1$, 
{\rm Pengyang Huang}$^1$, \\
{\rm \text{Junhan Pu}}$^1$, 
{\rm \text{Yuxin Cao}}$^3$, 
{\rm \text{Kai Ye}}$^4$, 
{\rm \text{Jie Hao}}$^1$ \mkletter \rm, and
{\rm \text{Yixian Yang}}$^1$
\thanks{\mkletter~Corresponding authors: \texttt{derek.wang@data61.csiro.au}, \\ \texttt{haojie@bupt.edu.cn}.}
\\
\small $^1$Beijing University of Posts and Telecommunications \ \
$^2$CSIRO's Data61 \\
\small $^3$National University of Singapore \ \
$^4$The University of Hong Kong \\
}

\maketitle

\begin{abstract}
  Speech synthesis technology has brought great convenience, while the widespread usage of realistic deepfake audio has triggered hazards. Malicious adversaries may unauthorizedly collect victims' speeches and clone a similar voice for illegal exploitation (\textit{e.g.}, telecom fraud). 
  However, the existing defense methods cannot effectively prevent deepfake exploitation and are vulnerable to robust training techniques. Therefore, a more effective and robust data protection method is urgently needed. 
  In response, we propose a defensive framework, \textit{\textbf{SafeSpeech}}, which protects the users' audio before uploading by embedding imperceptible perturbations on original speeches to prevent high-quality synthetic speech. In SafeSpeech, we devise a robust and universal proactive protection technique, \textbf{S}peech \textbf{PE}rturbative \textbf{C}oncealment (\textbf{SPEC}), that leverages a surrogate model to generate universally applicable perturbation for generative synthetic models. Moreover, we optimize the human perception of embedded perturbation in terms of time and frequency domains.
  To evaluate our method comprehensively, we conduct extensive experiments across advanced models and datasets, both subjectively and objectively. Our experimental results demonstrate that SafeSpeech achieves state-of-the-art (SOTA) voice protection effectiveness and transferability and is highly robust against advanced adaptive adversaries. Moreover, SafeSpeech has real-time capability in real-world tests. 
  The source code is available at \href{https://github.com/wxzyd123/SafeSpeech}{https://github.com/wxzyd123/SafeSpeech}.
\end{abstract}

\section{Introduction}\label{section_intro}

\begin{figure}[t]
    \centering
    \includegraphics[width=0.75\linewidth]{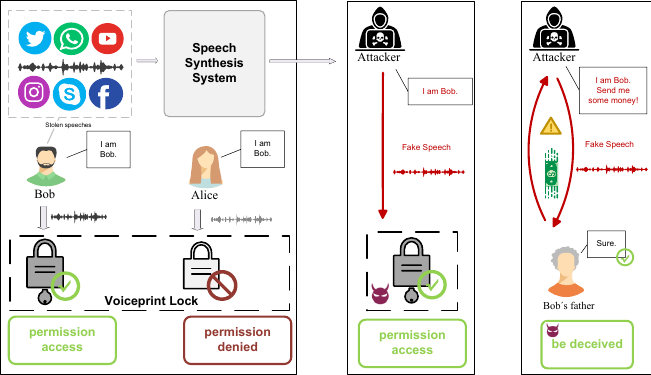}
    \caption{Speech synthesis hazards in real-world scenarios, \textit{e.g.}, the attacker utilizes Bob's public voices with TTS tools to bypass the voiceprint lock and achieve telecom fraud.}
    \label{fig_synthesis_attack}
\end{figure}

In recent years, the rapid growth of generative artificial intelligence (AI)~\cite{achiam2023gpt} has drawn broad social attention. People are amazed by the excellent capabilities of AI, which benefits from the continuous progress in deep neural networks (DNNs). 
In speech synthesis, or voice cloning, models trained on large-scale speech corpus can now generate highly realistic audio~\cite{kim2021conditional, kim2020glow, kawamura2023lightweight}. Through fine-tuning pre-trained models, the latest one needs only a few minutes of speech samples to synthesize high-quality speeches with realistic timbre, rhythm, and phonemes. Although early voice cloning technology is mostly used for positive purposes, such as cloning deceased lovers' voices to provide comfort, there have recently been cases of this tool being abused for illegal activities, \textit{e.g.}, Figure~\ref{fig_synthesis_attack}.
Moreover, criminals used deepfake speech to pose as a German boss and tricked a British subsidiary head into transferring \$243,000~\cite{stupp2019fraudsters}. Tackling deepfake speech is vital for the integrity and security of voice-based systems in daily life.

\noindent\textbf{Existing Defenses and Limitations.}
To counter the threat of deepfake speech, existing voice protection methods like AntiFake~\cite{yu2023antifake}, VSMask~\cite{wang2023vsmask}, and AttackVC~\cite{huang2021defending}, focus on leveraging adversarial examples to make the synthetic samples do not resemble the original speaker in terms of timbre, preventing zero-shot speech synthesis (or voice conversion).

However, although previous methods have certain effects, they also have serious limitations: 
(1) {\it \uline{Protection Scenarios}}. Current voice protection techniques focus on zero-shot scenarios, \textit{i.e.}, employing one reference audio to clone voice during the inference stage. However, in addition to zero-shot scenarios, adversaries may also fine-tune models, which poses a more severe challenge for two reasons. Firstly, many models do not support zero-shot systhesis~\cite{kim2021conditional, kim2020glow,kawamura2023lightweight}, and fine-tuning can achieve better quality. Secondly, previous methods based on adversarial examples cannot withstand fine-tuning techniques. 
(2) {\it \uline{Synthesis Quality Prevention}}. 
Previous voice protection methods on adversarial examples can generate dissimilar but high-quality deepfake speeches, which means that these speeches can still be utilized. 
However, we aim to make the synthesized audio significantly low-quality and cannot be utilized to address the deepfake issue fundamentally, \textit{i.e.}, \textit{synthesis quality prevention}. High-quality deepfake speeches pose security risks.
First, adversaries can conduct large-scale searches for new victims with similar voices. Second, synthetic audio might still be utilized maliciously, such as voice assistants in telecom fraud, to conceal the true identity.

\noindent\textbf{Motivation.} Many regions have introduced regulations on generative AI governance and data protection, like California's Consumer Privacy Act (CCPA)~\cite{ccpa2018}, making voice privacy protection urgent. We conduct this work for two motivations. Firstly, we aim for broader voice protection, covering training time and deepfake audio quality. Fine-tuning-based speech synthesis is crucial as it can cover more TTS models and produce higher-quality audio. Secondly, large language models (LLMs)~\cite{touvron2023llama2, achiam2023gpt} have developed continuously. LLMs can generate high-quality human-like text for TTS models, promoting realistic deepfake speech production. Synthesizers incorporating LLMs (\textit{e.g.}, BERT-VITS2~\cite{github:Bert-VITS2} and GPT-SoVITS~\cite{github:GPT-SoVITS}) also need consideration.

\noindent\textbf{Technical Challenges.} In addressing the aforementioned issues, we have to overcome these challenges: (1) {\it \uline{Effectiveness and Transferability}}. We need to design voice protection that is effective against fine-tuning. Additionally, the algorithm should possess strong transferability across various TTS models. (2)  {\it \uline{Modal Selection}}. The input of TTS models is multimodal, such as waveform, spectrogram, and corresponding text. It is crucial to decide the most sensible modality of the anti-learning perturbations.
(3)  {\it \uline{Robustness.}} Previous data protection methods~\cite{huang2021unlearnable, yu2023antifake} are vulnerable to adaptive training~\cite{wu2023onepixel}. Therefore, our method should be robust against advanced adaptive adversaries for real-world applications. (4)  {\it \uline{Imperceptibility.}} The embedded perturbation should be imperceptible or align with human perception, necessitating a design optimization method for noise incorporation that ensures human acceptance or harmlessness. 
Overall, successful voice protection should satisfy the prevention of synthetic speech intelligibility (\textit{i.e.}, synthesis quality) and speaker timbre similarity against training-stage voice cloning.

\noindent\textbf{Our Response Strategies.~} 
In response to these challenges, we introduce \textbf{\textit{SafeSpeech}}, a framework to safeguard data by embedding specifically designed perturbation while preserving text consistency before audio uploading. To effectively protect voice at training time and enhance transferability, we propose pivotal objective optimization with less computational time based on a surrogate model. Additionally, to achieve further protection, we introduce the Speech PErturbative Concealment (SPEC) techniques based on Kullback-Leibler divergence, which better conceals speech information. These approaches lead to voice protection in terms of speech quality and timbre similarity.  
To optimize the audibility of embedded noise, $\ell_p$ norm may not fully adapt to human~\cite{yu2023antifake} and we devise perceptual optimization functions to reduce human audibility.  
The safeguarded audio by our proposed SafeSpeech ensures that the synthesized audio is not similar and undermines the speeches' usability, thus providing a more effective and robust defense against various adaptive attackers utilizing a novel voice protection method.

\begin{figure*}[t]
    \centerline{
    \includegraphics[width=0.88\textwidth]{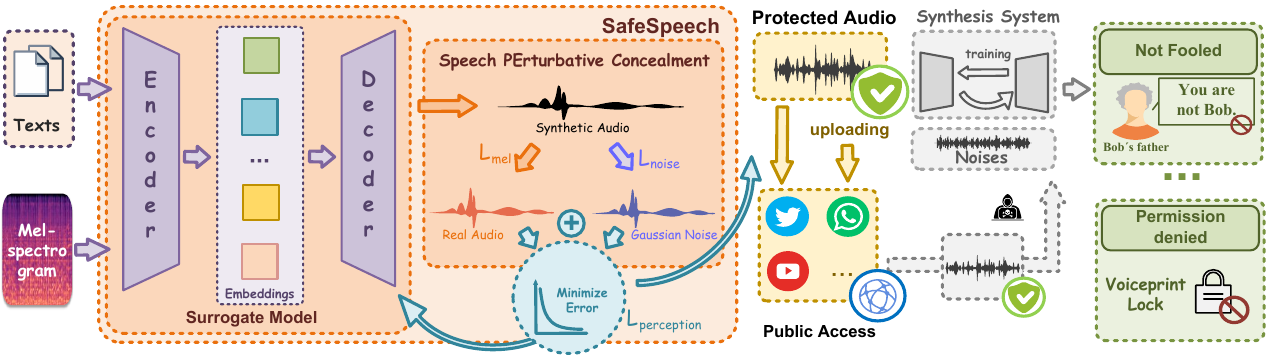}}
    \caption{The SafeSpeech safeguards voice by constructing a surrogate TTS model that minimizes the designed objectives ($\mathcal{L}_{mel}$ and $\mathcal{L}_{noise}$ with perception constraint $\mathcal{L}_{perception}$ detailed in Section \ref{section_method}). Despite attackers fine-tuning advanced TTS models from social platforms, they cannot produce high-quality synthetic speech to circumvent voiceprint locks or deceive victims' families.}
    \label{fig_workflow}
\end{figure*}

\noindent\textbf{Experiments and Evaluation.~} 
To validate the effectiveness of SafeSpeech, we conduct experiments on current SOTA TTS models and well-known datasets across various metrics. Our extensive experiments demonstrate that SafeSpeech achieves the SOTA protection effect against train/test-time voice cloning. 
SafeSpeech is also highly robust facing perturbation removal, data augmentation, model (or data) recovery, and adaptive robust training, \textit{etc}. Moreover, the physical-world test demonstrates that SafeSpeech possesses high robustness and real-time capabilities with continuous on-site voice protection. In the user study, we evaluate the human perception and most participants believe synthetic speeches after protection cannot deceive them. Compared to previous methods~\cite{yu2023antifake,wang2023vsmask,huang2021defending}, SafeSpeech has stronger effectiveness and robustness, preventing fine-tuning-based speech synthesis.

\noindent\textbf{Novelty and Contributions.} 
In this paper, we achieve innovation in three levels: (1) {\it \uline{Algorithm}}. We propose the pivotal objective optimization and the SPEC technique, which can achieve better effect, transferability, robustness, and efficiency. (2) {\it \uline{ Scenario}}. Compared to previous literature~\cite{yu2023antifake,wang2023vsmask,huang2021defending}, we consider the fine-tuning strategies, presenting a more protective scenario than zero-shot voice cloning. (3) {\it \uline{ Application}}. By utilizing a lightweight surrogate model and simplified objective, we can achieve real-time protection in real-world applications.
The CCPA emphasizes personal data privacy and regulates business data handling. Our SafeSpeech prevents unauthorized and malicious voice exploitation.
Our main contributions can be summarized as follows:

\begin{itemize}[leftmargin=1em, topsep=0pt, itemsep=0pt, parsep=0pt]
    \item We propose \textbf{\textit{SafeSpeech}} which for the first time protects our voice at training time in our best knowledge by embedding imperceptible perturbation against unauthorized exploitation and malicious speech synthesis.
    \item We devise a robust and universal perturbative technique named Speech PErturbative Concealment against malicious speech synthesis. For noise imperceptibility, we introduce a hybrid perceptual function, combining STOI and STFT loss, to optimize human perception and reduce inaudibility in terms of time and frequency domains. 
    \item We comprehensively evaluate SafeSpeech across ten SOTA models and two datasets during training and testing phases. The SafeSpeech is robust against adaptive adversaries.
    \item SafeSpeech can achieve real-time protection in our real-world test and takes only 10.606 seconds to generate speaker-specific perturbation with continuous protection. 
\end{itemize}

\section{Preliminaries}\label{section_preliminaries}
Mainstream speech synthesis utilizes a DNN-based model to input signals with timbre and audio features, and these models usually consist of an encoder and decoder architecture (\textit{e.g.}, Figure \ref{fig_workflow}). Compared to traditional rule-based synthesis methods~\cite{ning2019review}, current TTS models~\cite{kim2021conditional, kawamura2023lightweight, github:Bert-VITS2} can achieve a better voice cloning effect with a few samples. In this context, we explore privacy preservation strategies for TTS synthesis.  

\noindent\textbf{Voice Anti-Cloning.~}AntiFake~\cite{yu2023antifake}, VSMask~\cite{wang2023vsmask}, and AttackVC~\cite{huang2021defending} are three voice protection methods based on adversarial examples ensuring zero-shot TTS models cannot synthesize voiceprint-similar speech. Our SafeSpeech, in contrast, goes beyond mere voiceprint similarity and inference stage protection, as it actively prevents the usability of deepfake speech. SafeSpeech considers the protection of the timbre feature and synthesis quality \textit{based on unlearnable examples, a training stage data protection technique}. 
%Therefore, we protect voices on a much broader level.

\noindent\textbf{Data Poisoning.~} Our method can be regarded as a special type of clean-label and triggerless data poisoning, while the task purposes are significantly distinct. Data poisoning attack aims to degrade the model's performance on clean samples by modifying training samples. Previous data poisoning attacks~\cite{munoz2017towards, yang2017generative} have focused on identifying the most influential samples to affect model learning and modifying these training samples (\textit{e.g.}, changing labels~\cite{tolpegin2020data} or embedding large perturbations~\cite{yang2017generative}). The primary aim of poisoning attacks is to degrade the overall performance of the model after poisoning a portion of the data resulting in the inability to use the authorized data normally, while in the protective scenario discussed in this paper, we aim to protect audio data rather than poisoning the model to degrade its performance. In other words, the perturbations added by SafeSpeech affect the protected samples without affecting the unprotected ones.  This point will be discussed in Section \ref{section_robustness_finetunig} by experiments.

\noindent\textbf{Unlearnable Examples.~} In classification tasks, let a DNN-based classifier be $f$, which accepts input data $x$ with corresponding label $y$. We regard the clean training and testing datasets as $D_c$ and $D_t$, respectively. The creation of unlearnable examples is facilitated by a perturbation optimizer that treats $D_c$ as the input and produces an unlearnable dataset $D_u$ by embedding a perturbation $\delta$ to samples in $D_c$. When training on $D_c$, the model $f$ demonstrates excellent performance on $D_t$ but suffers from poor performance when training on $D_u$. The perturbation generation relies on a bi-level structure~\cite{huang2021unlearnable} which optimizes both the perturbation and the parameter:
\begin{equation}
\argmin \limits_{\theta} \mathbb{E}_{x,y}[\min_{\delta} \mathcal{L}(f_\theta(x+\delta), y)],\label{eq_unleanrable}
\end{equation}
where $f_\theta$ is a classifier with the trainable parameter $\theta$.

Research~\cite{huang2021unlearnable, fu2022robust, zhang2024mitigating} indicates that minimizing errors in the training objectives can initially disrupt the training process. However, to achieve effective and robust voice protection, more advanced techniques must be designed.

\section{Threat Models}\label{section_threat_model}
In the threat model, we introduce 
%entities' description, including 
\textit{a priori} knowledge, capability, and limitations of the adversary, defender, and system.

\subsection{Adversary Capability}
We assume that the adversary is the third-party entity that can utilize the current most advanced TTS models to achieve successful voice cloning of the victim on unprotected data. To simulate experienced adversaries in the real world, we consider three of their capabilities:

\noindent\textbf{Capability of Data Access.} The development of the Internet has exposed more data to the public with potential threats. Adversaries can directly download the uploaded audio of the victim from public media on the Internet (\textit{e.g.}, YouTube, Facebook) using web crawler technology or some permission bypass mechanisms. They cannot access the original unprotected audio if obtained speeches are protected.

\noindent\textbf{Capability of Customized and Robust Training.} Adversaries can leverage various advanced speech synthesis models. Attackers can achieve speech synthesis to bypass speaker verification or human perception. At the same time, we consider a stronger adversary, that is, the adversary can detect the abnormal perturbations embedded for protection and employs the most advanced defensive data augmentation and robustness training, such as perturbation removal, adversarial training, specific data poisoning defensive methods, and speech transformation, to seek high-quality speech synthesis.

\noindent\textbf{Capability of Model Recovery.} We assume that the adversary is an experienced model trainer. If the speech synthesis does not achieve the desired effect after fine-tuning the TTS model utilizing the acquired audio, then he may realize that the model has been poisoned and restore it to the initial state against the protection strength detailed in Section \ref{section_robust_recovery}.

\subsection{Defender Description}\label{section_threat_defender}
\noindent\textbf{Defender Limitations.~} To more faithfully replicate real-world scenarios, we restrict the defender to only having access to the synthesized audio from the model's output and is unaware of the model training method of the attacker. The defender can only introduce perturbation to the original audio. Additionally, we restrict that the generation of noise does not depend on \textit{a priori} knowledge for users' application, \textit{e.g.}, the utilized model for speech synthesis, and the protection of the speaker's samples can poison all or only part of the samples. Moreover, we assume the uploading condition when the protected audio can defend potential and the most advanced data augmentation and robust training, \textit{etc.}, strategies for confidential exploitation in the real world after uploading the perturbative protected samples.

\noindent\textbf{System Capabilities.~} System, \textit{i.e.}, SafeSpeech in this paper, needs to effectively generate perturbation for the audio that users want to protect before uploading. For the aim of voice protection, SafeSpeech should generate perturbation without affecting the use of unprotected audio. Moreover, the generated perturbation should be imperceptible or at least perceptually acceptable. Audio protected by SafeSpeech should achieve timbre and synthesis quality protection, while also ensuring transferability across different models.
In conclusion, we design our protective system for the two aims:

\begin{itemize}[leftmargin=1em, topsep=0pt, itemsep=0pt, parsep=0pt]
    \item \textbf{System Aim 1 (SA1): Timbre Protection.} Our goal is to achieve a state where fine-tuning on protected audio cannot synthesize audio that resembles the target victim.
    \item \textbf{System Aim 2 (SA2): Quality Protection.} The ``synthesis'' quality protection represents the synthetic speech is low-quality and cannot be utilized normally in daily life.
\end{itemize}

Previous adversarial-examples-based voice protection~\cite{yu2023antifake, wang2023vsmask,huang2021defending} can only achieve SA1 in zero-shot scenarios. However, to fundamentally prevent deepfake audio, the defender needs to meet SA1 and SA2 in zero-shot and fine-tuning scenarios.

\section{SafeSpeech Methodology}\label{section_method}

In response to speech synthesis defense at training time, we design SafeSpeech to achieve voice anti-cloning. Figure \ref{fig_workflow} presents the workflow of SafeSpeech and the attacker's malicious action. For data protection, we introduce the optimization objective and propose the proactive defense mechanisms of the pivotal function and SPEC. While, for perception optimization, we introduce the perception metrics, \textit{i.e.}, STFT, and STOI, to better the human perception of protected samples.

\noindent\textbf{Problem Formulation.~} As we introduced in Section \ref{section_intro}, we aim to prevent high-quality deepfake audio generation and propose a universal and robust perturbative voice protection method. Based on this, we design to solve an error-minimizing problem including effect and perception. We express the objectives of SafeSpeech by the following formula:
\begin{equation}\label{eq_problem}
    \begin{aligned}
        \argmin_{\delta}\ & \mathcal{L}(G(x+\delta), x) + \alpha P(x+\delta), \\
        s.t.\ & H(G(x+\delta)) \ne H(G(x)),\\
              & SV(G(x+\delta)) \ne SV(G(x)),\\
              & H(x+\delta) \approx H(x),
    \end{aligned}
\end{equation}
where $G(\cdot)$, $SV(\cdot)$ are a speech synthesizer and speaker verification system respectively, $P(\cdot)$ is the auditory function, and $H(\cdot)$ is the human perception according to input audio. $\mathcal{L}(\cdot)$ is the objective, $x$ is an input audio, $\alpha$ is a weight coefficient, and $\delta$ is the perturbation bounded by $\ell_p$ norm as $||\delta||_p \le$ radius $\epsilon$ for the limitation of human perception.

\subsection{Data Protection }\label{section_method_spec}
To effectively mitigate the unauthorized speech synthesis in real-world scenarios, the generation process of protected audio should be devoid of reliance on \textit{a priori} knowledge. This is crucial as we cannot predict the training strategies of the adaptive attackers. In the design of SafeSpeech, we aim to produce an effective, robust, and universal perturbation preventing training-time speech synthesis across different models.

\noindent\textbf{Unlearnable Audio}. 
Previous voice protection based on adversarial examples cannot be effective during the training stage, which is an unavoidable scenario. We aim to solve this based on training-stage data protection~\cite{huang2021unlearnable} and introduce an {\it error-minimizing} (EM) noise. The EM problem reduces the error of the model simulating the normal training process by perturbation so that there is ``nothing'' to learn when training on the safeguarded dataset as we introduced in Eq. (\ref{eq_unleanrable}).

It is significant to decide the optimization objective, because TTS models often engage in multi-task learning with multimodal inputs, like audio accompanied with relevant text guidance. Under the assumption of the defender, we can only modify the user's original audio while preserving the integrity of the text input. Generative speech synthesis models typically learn from the input data and generate outputs with similar distributions. They may also incorporate discriminators or speaker encoders to improve the high-quality audio generation. However, considering the various architectures of different models, we focus on optimizing the generator, as it is most related to output information and audio quality.

By optimizing the objective function using perturbation, the error of the model has been greatly reduced and it is possible to make audio unlearnable for the model to think that there is ``nothing'' to learn. The objective function of the generator $g$ from a TTS model can be expressed by following the single-level loop with multi-modal inputs:
\begin{equation}
    \mathcal{L}_{TTS} = \sum\limits_{i=0}^{k} \mathcal{L}_i\left[g(\texttt{text}, \texttt{spec}(x+\delta)), x, \theta \right],\label{eq6}
\end{equation}
where $x$ represents the raw waveform, and $\text{spec}(\cdot)$ computes the linear spectrogram from inputs, and $\theta$ is the parameter.

Therefore, the core solution is to decide $\mathcal{L}_i$.

\noindent\textbf{Pivotal Objective Selection}.
When applying noise to optimize the multi-task learning problem, directly using the objectives of TTS models yields unsatisfactory perturbative results (Section \ref{section_exp_ablation}). Moreover, this is highly inefficient, requiring specifying a unique optimization approach for each TTS model. In this part, we will first illustrate why direct optimization is ineffective through examples, and then proceed to carefully analyze the features of TTS models and propose our pivotal objective function and its principles to be satisfied.

Different TTS models own different objective functions and components. For instance, BERT-VITS2~\cite{github:Bert-VITS2} comprises four components, with the generator containing eight objective functions, and the optimization function can be expressed as:
\begin{dmath}    \mathcal{L}_{G}=\mathcal{L}_{recon}+\mathcal{L}_{kl}+\mathcal{L}_{dur}+\mathcal{L}_{adv}(G)+\mathcal{L}_{fm}(G)+\mathcal{L}_{DurD}
    +\mathcal{L}_{score}+\mathcal{L}_{encoder}, \label{eq_bert}
\end{dmath}
where $\mathcal{L}_{recon}$ denotes the reconstruction loss between ground truth and generated speech. $\mathcal{L}_{kl}$ and $\mathcal{L}_{dur}$ represent the KL divergence loss and duration loss. $\mathcal{L}_{adv}$ and $\mathcal{L}_{fm}$ are the adversarial training loss and feature-matching loss of the generator. $\mathcal{L}_{DurD}$ is the duration discriminator loss, $\mathcal{L}_{score}$ computes the similarity score of embeddings from the generated and real audio, and $\mathcal{L}_{encoder}$ represents the encoder loss.

Among these, the duration loss $\mathcal{L}_{dur}$ only depends on the input text and cannot be optimized via perturbation. Furthermore, VITS~\cite{kim2021conditional} has five objective functions that are different from BERT-VITS2. Consequently, directly applying the eight-loss setup from BERT-VITS2 for perturbation may not yield a universally applicable approach in VITS or other models.
 
Moreover, through a thorough analysis of the objective function, we realize that the optimization effect is closely related to the multi-modal input characteristics of the model. We can only interfere with audio waveform, so that objective functions unrelated to audio can not be effectively affected. Furthermore, due to the various structures and optimization objectives of different models, it is advisable to devise a \underline{\it universal} function adapting different generative TTS models for the perturbation universality.

The selection of an optimization objective is crucial given our lack of knowledge about the attacker's training model and structure. So, we should stick to the following  listed principles of the pivotal function selection: 
\begin{enumerate}[leftmargin=2em, topsep=0pt, itemsep=0pt, parsep=0pt]
    \item[(a)] The objective function can be optimized by perturbation;
    \item[(b)] To ensure that universal perturbation is independent of \textit{a priori} knowledge, the objective function should be universal across various TTS models;
    \item[(c)] The function should be easily optimized through perturbation, such as achieving a rapid rate of convergence or containing relatively rich information entropy.
\end{enumerate}

For example, in Eq. (\ref{eq_bert}), the $\mathcal{L}_{dur}$ is not related to the waveform $x$ violating the principle (a). These three principles must be considered in data protection guiding the transferability of the perturbation. When designing SafeSpeech, it is crucial to ensure that the users can perform perturbations regardless of the model they are using. This poses a challenging question: 

\begin{tcolorbox}
  \begin{center}
    \textit{Is it possible to devise a perturbative method that is universally applicable for all generative TTS models?}
  \end{center}
\end{tcolorbox}

In the scenario of this paper, a universal perturbation method implies identifying an optimization target that is consistent across different models. Upon further reflection, we recognize that generative TTS models typically aim to output audio or spectrograms that follow a distribution similar to the input~\cite{kim2021conditional, kim2020glow, kawamura2023lightweight, ren2021fastspeech, shen2018natural, li2019neural, tan2024naturalspeech, shen2024naturalspeech,li2024styletts}, with training involving fitting these distributions to optimize the generator. Therefore, we propose to measure the {\it \uline {distance between the model's output and the input waveform as our pivotal optimization target}}. Drawing from common TTS objective functions, we select the $\ell_1$ distance to compute the similarity between the mel-spectrograms of the synthesized audio $\hat{x}$ and protected real audio $x'$. For end-to-end TTS, we first compute the mel-spectrograms for both the output and the input audio. If the model outputs spectrograms, we can then directly optimize them. This mel function can be formulated as:
\begin{equation}
    \mathcal{L}_{mel} = ||x'_{mel} - \hat{x}_{mel}||_1 .\label{eq_mel}
\end{equation}

\begin{figure}[t]
\centerline{\includegraphics[width=0.3\textwidth]{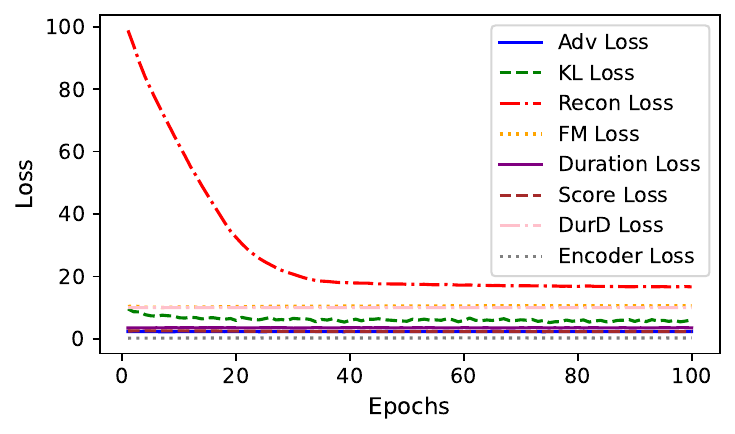}}
\caption{The convergence speed comparison of different objective functions when optimizing by perturbation.}
\label{fig_convergence}
\end{figure}

The reason for choosing the function $\mathcal{L}_{mel}$ as our pivotal objective lies in the fact that generative TTS models serve the synthetic speeches $\hat{x}$ as the models' outputs. Crucially, we possess the capability to calculate the $\ell_1$ distance between these outputs and real speeches, a computation that remains agnostic to the models' complex architecture. The \texttt{mel} optimization function $\mathcal{L}_{mel}$ also satisfies the principles (a) and (b). In the following, we analyze the selection from the perspective of convergence speed of principle (c). We use 10 samples for a batch and optimize them for 100 epochs on BERT-VITS2. The convergence speed is presented in Figure \ref{fig_convergence}. We can observe that the \texttt{recon} function $\mathcal{L}_{mel}$ has the fastest convergence speed, starting from the highest initial value of 98.8 and eventually dropping to 16.7, and approaches convergence at the 40th epoch. In contrast, other objective functions have slow convergence speeds with hardly any noticeable decrease, so they are hard to optimize via perturbation. Therefore, we target the objective function with the fastest convergence speed and easiest optimization as a part of the data protection.

Compared with vanilla objective optimization, \textit{i.e.}, initial unlearnable examples~\cite{huang2021unlearnable}, selecting a generalized and fastest-converging function from multi-objective functions for optimization can enhance the effectiveness of protection and reduce time costs. This is because, when using perturbation to optimize the multi-task objectives, the values in the gradient space are influenced by the optimization directions of multiple functions, making it difficult for the loss to converge. In contrast, choosing $\mathcal{L}_{mel}$ allows the gradient to be optimized only in one and the fastest direction, thereby enabling a universal and effective protection method.

The experimental comparison between utilizing vanilla and simplified optimization objective is presented in Section \ref{section_exp_ablation}. Drawing from the aforementioned analyses, we select the pivotal objective $\mathcal{L}_{mel}$ as a core component of SafeSpeech.

\noindent\textbf{Speech PErturbative Concealment}. 
During the optimization process, we find that optimizing Eq. (\ref{eq_mel}) can effectively make the synthesizer produce unclear speech. However, if we listen attentively, there are still parts of the slightly auditory pronunciation, and its background noise is relatively low. Moreover, we hope that the synthesizer can only generate noise when training on a SafeSpeech-protected dataset to achieve SA1 and SA2, and the robustness of the perturbation must be guaranteed. Based on this, we aim to propose a more effective and robust defense mechanism. When training the synthesizer, we expect the generator to produce noise. Therefore, in the perturbation optimization procedure, we utilize a Gaussian-distributed random noise $z$, leading the generator to produce noise. 
We aim that the distribution of the synthesized output can increasingly approximate a real noise distribution $z$. Based on this, we employ the Kullback-Leibler (KL) divergence as part of the optimization objective due to its \underline{\it asymmetry}.
Lower values of KL scatter mean that the output of the model is more similar to the noise distribution, which can achieve the low-quality of deepfake speech.
At the same time, for both the random noise and generated audio, we extract mel-spectrogram feature $z_{mel}$ and $\hat{x}_{mel}$, to acquire more relative information, and reduce the $\ell_1$ distance as Eq. (\ref{eq_mel}). The objective function can be described as:
\begin{equation}
    % \mathcal{L}_{noise} =  \sum p(\hat{x}_{mel})\left[\log p(\hat{x}_{mel}) - \log q(z_{mel})  \right] +  ||\hat{x}_{mel} - z_{mel}||_1.\label{eq_noise}
    \mathcal{L}_{noise} =  D_{KL}(\hat{x}_{mel}, z_{mel}) +  ||\hat{x}_{mel} - z_{mel}||_1, \label{eq_noise}
\end{equation}
where $D_{KL}$ represents the KL divergence of two distributions.

Based on the above, we desire the model to learn the perturbations rather than speech information. To achieve this, we propose a method named Speech PErturbative Concealment (SPEC), which combines the mel function and the noise loss function assigning a suitable weight and can be expressed by:
\begin{equation}
    \mathcal{L}_{SPEC} = \mathcal{L}_{mel} + \beta \mathcal{L}_{noise}.\label{eq_SPEC}
\end{equation}
where $\beta$ is the hyperparameter to be set.

In conclusion, we introduce a pivotal objective function aimed at simplifying the multi-task learning problem to achieve effectiveness at the training stage and enhance transferability. Furthermore, we design the Speech PErturbative Concealment method that measures the two distributions between synthesized speech and random noise, thereby concealing the speaker's information with a stronger protection.

\subsection{Perception Optimization }\label{section_method_perception}
The perturbation should be generated without interfering with the normal exploitation of data samples. So the imperceptibility of the noise is an important factor. In the process of noise generation introduced in Section \ref{section_method_spec}, $\ell_p$ norm is employed to limit the perturbation boundary so that the overall magnitude of the noise values can not be particularly large. However, the limitation of the perturbation in the value aspect cannot completely represent human perception. Based on this, SafeSpeech utilizes the noise perception module to reduce the gap between the $\ell_p$ norm and human perception. We optimize the audio perception in the time and frequency domains. 

For better noise perception and imperceptibility, we employ Short-Time Objective Intelligibility (STOI)~\cite{zhang2018training} score as our main part of the perception module. STOI score represents speech intelligibility as an objective metric, which computes the correlation of short-time temporal envelopes of the clean and protected audio, ranging from 0 to 1 and a higher score indicates better speech quality. STOI score is closely related to the human auditory perception and optimizing STOI function $\mathcal{L}_{stoi}$ brings a more natural sound. Moreover, we follow the principles to compute $\mathcal{L}_{stoi}$ introduced in~\cite{zhang2018training}.

On the other hand, we consider the time and frequency domain of audio for better perception optimization in the $\ell_p$ radius. Short-Time Fourier Transform (STFT)~\cite{HiddenSpeaker} performs well in feature extraction, so we utilize the $\ell_2$ distance as part of our perception loss which can be expressed by:
\begin{equation}
    \mathcal{L}_{stft} = ||\text{STFT}(x+\delta) - \text{STFT}(x)||_2.\label{eq_stft}
\end{equation}

Based on the above, the perception module of SafeSpeech crafts a hybrid optimization function:
\begin{equation}
    \mathcal{L}_{perception} = \mathcal{L}_{stoi} + \mathcal{L}_{stft}. \label{eq_perception}
\end{equation}

\noindent\textbf{Method Conclusion.} Combining proposed data protection and perception optimization techniques, the objectives of SafeSpeech $\mathcal{L}$ can be expressed by:
\begin{equation}
    \mathcal{L} = \mathcal{L}_{SPEC} + \alpha \mathcal{L}_{perception}, \label{eq_protection}
\end{equation}
where $\alpha$ is the weight coefficient the same as in Eq. (\ref{eq_problem}), which balances the imperceptibility and effectiveness in Section \ref{section_exp_ablation}.

Algorithm \ref{algori} shows the detailed description of SafeSpeech. The SafeSpeech can optimize the perturbation for $max\_epoch$ steps. \texttt{init_perturbation_set} assigns the perturbation to a random initial value within radius $\epsilon$. If the effectiveness performance dissatisfies the user's expectation, the hyperparameter, such as $max\_epoch$ and $\epsilon$, can be changed to enhance the protection performance until achieving a satisfactory level in the last step \texttt{evaluation\_optimize\_hyperparameters}.

\normalem
\begin{algorithm}[t]
\small
\SetAlgoLined
\textbf{Inputs}: $(x, text) \in D_c$, perturbation $\delta$, surrogate model $\mathcal{M}$, optimization numbers $max\_epoch$. \\
\textbf{Parameters}: random Gaussian noise $z$, $\ell_p$ norm boundary radius $\epsilon$, weight coefficients $\alpha$ and $\beta$. \\
\textbf{Output}: protected audio $x'$. \\
\nlset{1} $\delta \leftarrow \texttt{init\_perturbation\_set}(-\epsilon, \epsilon)$; \\
\nlset{2} $x' \leftarrow x + \delta$; \\
\nlset{3} \For{$j \leftarrow 1$ \KwTo $max\_epoch$}{
    \nlset{4} $\hat{x} \leftarrow \mathcal{M}(\texttt{spec}(x'), text, \texttt{other\_input})$; \\
    \nlset{5} $\mathcal{C}_1 \leftarrow \mathcal{L}_{mel}(\hat{x}_{mel}, x')$; \\
    \nlset{6} $\mathcal{C}_2 \leftarrow D_{KL}(\hat{x}_{mel}, z_{mel}) + \|\hat{x}_{mel} - z_{mel}\|_1$; \\
    \nlset{7}\eIf{$\texttt{Perception\_Optimize}$}{
        \nlset{8} $\mathcal{C}_3 \leftarrow \mathcal{L}_{perception}(x, x')$; \\
        \nlset{9} $\mathcal{C} \leftarrow \mathcal{C}_1 + \beta \cdot \mathcal{C}_2 + \alpha \cdot \mathcal{C}_3$; \\
    }
    {
        \nlset{10} $\mathcal{C} \leftarrow \mathcal{C}_1 + \beta \cdot \mathcal{C}_2$; \\
    }
    \nlset{11} $\delta \leftarrow \texttt{Clamp}(-\texttt{sign}(\nabla_x \mathcal{C}), -\epsilon, \epsilon)$; \\
    \nlset{12} $x' \leftarrow x + \delta$; \\
}
\nlset{13} \texttt{evaluation\_optimize\_hyperparameters}(). \\
\caption{SafeSpeech.}
\label{algori}
\end{algorithm}

To summarize, SafeSpeech achieves training-stage voice protection by introducing the pivotal objective and SPEC technique. In the pivotal objective, we innovatively select the function with the {\it \uline{fastest convergence rate and universality}} to optimize, \textit{i.e.} the $\mathcal{L}_{mel}$ in Eq. (\ref{eq_mel}). In the SPEC technique, considering the \uline{\it asymmetry} of the KL divergence, which can better measure the difference between real and synthetic distributions, we propose a speech concealing technique based on KL divergence to enhance the effectiveness, which is also novel compared to previous methods.

\section{Experimental Settings}\label{section_settings}
In this section, we describe our experimental settings on models, datasets, hyperparameters, and metrics. All the experiments were conducted on one NVIDIA A800 GPU.

\subsection{Baselines}\label{section_settings_baselines}
For a broader comparison, we consider two types of data protection: \uline{\textit{perturbative availability poisons}} (PAP)~\cite{liu2023image, yu2022availability}, which protects data during the training stage, and {\it \uline{voice protection}} techniques. Referring~\cite{liu2023image}, we employ SOTA PAP baselines, including AdvPoison~\cite{fowl2021adversarial}, SEP~\cite{chen2023selfensemble}, and PTA~\cite{huang2021unlearnable}. In terms of voice protection, we utilize two open-source protection approaches: AntiFake~\cite{yu2023antifake} and AttackVC~\cite{huang2021defending}. We provide a detailed comparison in Appendix \ref{section_work_comparison}.

\noindent\textbf{Adversarial Poisoning (AdvPoison)}~\cite{fowl2021adversarial}. Fowl \textit{et al.}~\cite{fowl2021adversarial} demonstrated that adversarial examples, particularly targeted attacks, can achieve more protective effectiveness.

\noindent\textbf{Self-Ensemble Protection (SEP)}~\cite{chen2023selfensemble}. The perturbation dynamically interferes with a DNN during its whole training process. Based on this, Chen \textit{et al.}~\cite{chen2023selfensemble} proposed self-ensemble protection. It uses intermediate checkpoint models in a self-ensemble way to enhance and better simulate a training and dynamic model, improving perturbation generalization.

\noindent\textbf{Patch-To-All (PTA)}. For VITS~\cite{kim2021conditional}, MB-iSTFT-VITS~\cite{kawamura2023lightweight} and BERT-VITS2, a fast and efficient training strategy, windowed generator training, is utilized to randomly crop a fixed-length sample from a complete audio. Drawing inspiration from comparable process methods employed in adversarial attacks, we devise an approach, Patch-To-All, which generates perturbation that minimizes the error from fragment audio~\cite{huang2021unlearnable} and patches it to the entire sample as a comparison.

\noindent\textbf{AntiFake}~\cite{yu2023antifake}, \textbf{AttackVC}~\cite{huang2021defending}. Given a clean sample $x$ from speaker victim $i$, the speaker's timbre feature $E_i$ is computed by an encoder. Subsequently, a targeted speaker $j$ with the least similar timbre (in AntiFake) or randomly selected with the opposite gender (in AttackVC) is identified with timbre feature $E_j$. A perturbation is added to the original sample $x$ in such a way that $E_i$ becomes similar to $E_j$, thereby accomplishing voice cloning that results in timbre dissimilarity to the speaker victim $i$. We utilize the tools they have released to convert an original speech into a protected one.

\subsection{Text-To-Speech Synthesizers}\label{section_settings_models}
To comprehensively evaluate the effectiveness and transferability of SafeSpeech, we have selected a range of models for evaluation. On the one hand, we choose some classic, widely used, and improved models with fine-tuning capabilities in the TTS field. On the other hand, we select the latest and top-performing SOTA models based on the benchmark, TTSDS~\cite{ttsds}. These models vary in architecture, encompassing those based on {\it generative flow} (GlowTTS~\cite{kim2020glow}), {\it Variational Autoencoder} (VAE) architectures (VITS~\cite{kim2021conditional} and MB-iSTFT-VITS~\cite{kawamura2023lightweight}), {\it encoder-decoder} frameworks (OpenVoice~\cite{qin2023openvoice}), {\it diffusion models} (StyleTTS 2~\cite{li2024styletts} and TorToise-TTS~\cite{betker2023better}), and {\it flow matching} (F5-TTS~\cite{chen2024f5}).

Due to the remarkable capabilities of LLMs in dialogue and text generation, the advanced and latest TTS models, \textit{e.g.}, BERT-VITS2~\cite{github:Bert-VITS2}, XTTS~\cite{casanova2024xtts}, and FishSpeech~\cite{liao2024fish}, have generally integrated LLM components with synthesizers. In this setup, the LLM learns and emulates pronunciation characteristics and speaking styles of the target speakers, while the synthesizer focuses on learning and replicating the timbre features. This combination has led to a significant improvement in terms of synthetic naturalness. We provide a more detailed and comparative introduction to each model in Appendix \ref{sectionA}.

We utilize models with fine-tuning capabilities, \textit{i.e.}, BERT-VITS2, StyleTTS 2, MB-iSTFT-VITS, VITS, and GlowTTS, to validate the protective effect at training time in Section \ref{section_exp_finetuning}, and zero-shot models, \textit{i.e.}, TorToise-TTS, XTTS, OpenVoice, FishSpeech, and F5-TTS, to evaluate at inference time in Section \ref{section_exp_zero-shot}. For GlowTTS, VITS, MB-iSTFT-VITS, and StyleTTS2, we use pre-trained models on LJSpeech~\cite{ljspeech17} speech corpus. For BERT-VITS2, we utilize the model trained on a large-scale multilingual and multi-speaker speech corpus.

\subsection{Experimental Datasets}\label{section_settings_datasets}
We leverage utterances from two standard speech synthesis datasets for method assessment.
LibriTTS~\cite{zen2019libritts} is utilized to evaluate the performance of our method in targeted single-speaker with long sentences, while CMU ARCTIC~\cite{kominek2003cmu} is harnessed to assess multiple speakers with shorter sentences.

\noindent\textbf{LibriTTS}~\cite{zen2019libritts}. For effective fine-tuning, we select the top speaker who exhibits the highest similarity in voiceprint to the targeted speaker of LJSpeech.~\cite{ljspeech17} from the LibriTTS train-clean-100 subset which is derived from LibriSpeech\cite{panayotov2015librispeech} corpus, a large-scale of speakers corpus.

\noindent\textbf{CMU ARCTIC}~\cite{kominek2003cmu}. CMU ARCTIC includes audio recordings from 18 speakers and the speech content is nearly similar. For each speaker, we select 100 samples for fine-tuning.
For the training dataset, we randomly shuffled each dataset and employed 80\% of the audio samples for training, reserving the remaining 20\% for evaluation. More detailed information on the two datasets is presented in Appendix \ref{sectionA}.

\subsection{Hyperparameters and Metrics}\label{section_settings_metrics}
In this part, we outline the hyperparameters in our experiments and evaluation metrics objectively and subjectively.

\noindent\textbf{Hyperparameters.} In the fine-tuning process, we keep the conventional hyperparameters in~\cite{kim2021conditional, kim2020glow, kawamura2023lightweight, li2024styletts} while setting the correct sampling rate in our customized dataset. In noise generation, we set the perturbation radius $\epsilon$ as $8/255$ to reach a balance of human perceptibility and unlearnability and optimize the noise until the perturbation performs well. To ensure the effectiveness of synthesis, we train the models 100 epochs for single speaker and 200 epochs for multi-speaker datasets. It is worth noting that we changed the input of the models from spectrograms to the original waveform to better fit the realistic training scenarios. We set the $\alpha$ in Eq. (\ref{eq_protection}) as 0.05 to balance the imperceptibility and unlearnability and the $\beta$ in Eq. (\ref{eq_SPEC}) as set as 10 to achieve the best perturbative poisoning.

\noindent\textbf{Metrics}. We consider the subjective and objective metrics: 
\begin{itemize}[leftmargin=1em, topsep=0pt, itemsep=0pt, parsep=0pt]
    \item \underline{\it Mel-Cepstral Distortion} (MCD)~\cite{kubichek1993mel}: MCD, using Dynamic Time Warping mode, measures the disparity in audio features between synthesized and real audio, reflecting differences in speech content, timbre, \textit{etc}.
    \item \underline{\it Word Error Rate} (WER)~\cite{jiang2024megatts}: WER measures pronunciation clarity by a pre-trained medium-size Whisper~\cite{radford2023robust} to recognize text. {\it Higher MCD and WER represent worse speech clarity to achieve synthesis quality protection}.
    \item \underline{\it Speaker Similarity} (SIM)~\cite{jiang2024megatts}: SIM is a metric to evaluate the timbre similarity between two speeches. Higher SIM represents the larger timbre similarity. We follow the principles from \cite{jiang2024megatts} and leverage ECAPA-TDNN~\cite{desplanques20_interspeech} as the speaker encoder to compute the cosine similarity score between the real and synthetic speeches. 
    When the SIM exceeds 0.25, personal voice has been successfully cloned in timbre~\cite{desplanques20_interspeech}. The Attack Success Rate (ASR) is calculated by the ratio of successfully cloned samples in speaker similarity (\textit{i.e.}, SIM > 0.25) to the total number of samples.
    \item \underline{\it Signal-to-Noise Ratio} (SNR)~\cite{yu2023antifake}: SNR calculates the ratio between the embedded perturbation and the original audio to measure the levels of perturbation volume.
    \item \underline{\it Naturalness}~\cite{baba2024utmosv2}: We utilize the advanced DNN-based audio predictor, UTMOSv2~\cite{baba2024utmosv2} model, to evaluate the speech naturalness objectively. Moreover, we evaluate the naturalness subjective by human survey. {\it Higher values of SNR and Naturalness represent better speech quality.}
    \item \underline{\it Mean Opinion Score} (MOS)~\cite{kumar2019melgan}: MOS is a subjective evaluation metric that measures human perception of audio quality, typically ranging from 0 to 5, with higher values indicating better audio quality.
\end{itemize}

In conclusion, achieving higher MCD and WER values effectively fulfills SA2, preventing the malicious usage of synthesized samples. Simultaneously, maintaining a lower SIM adheres to SA1, realizing the identification protection.

\begin{table*}[t]
  \centering
  \caption{Comparison of the TTS models trained on clean, random Gaussian noise added, patch-to-all (PTA), adversarial poisoning (AdvPoison), self-ensemble protection (SEP), AntiFake, AttackVC, and our proposed Speech PErturbative Concealment (SPEC) safeguarded dataset. 
  The best and second-best unlearnability results are highlighted with \textbf{bold} and \underline{underlined}, respectively.}
  \resizebox{0.98\linewidth}{!}{
  \begin{threeparttable}
  \begin{tabular}{cccccc ccccc cccccc cccccc}
    \toprule
    \multirow{2}[1]{*}{Dataset} & \multirow{2}[1]{*}{Method}
    & \multicolumn{3}{c}{BERT-VITS2~\cite{github:Bert-VITS2}} 
    & \multicolumn{3}{c}{StyleTTS2~\cite{li2024styletts}}
    & \multicolumn{3}{c}{MB-iSTFT-VITS~\cite{kawamura2023lightweight}}
    & \multicolumn{3}{c}{VITS~\cite{kim2021conditional}} 
    & \multicolumn{3}{c}{GlowTTS~\cite{kim2020glow}}\\
    % \cmidrule(r){2-4}\cmidrule(lr){5-7}\cmidrule(lr){8-10}\cmidrule(l){11-3}
    \cmidrule(r){3-5}\cmidrule(lr){6-8}\cmidrule(lr){9-11}\cmidrule(lr){12-14}
    \cmidrule(lr){15-17}
    & & MCD($\uparrow$)   & WER(\%)($\uparrow$)  & SIM($\downarrow$)
    & MCD($\uparrow$)   & WER(\%)($\uparrow$)  & SIM($\downarrow$)
    & MCD($\uparrow$)   & WER(\%)($\uparrow$)  & SIM($\downarrow$)
    & MCD($\uparrow$)   & WER(\%)($\uparrow$)  & SIM($\downarrow$)
    & MCD($\uparrow$)   & WER(\%)($\uparrow$)  & SIM($\downarrow$)
    \\
    \midrule
    \multirow{9}{*}{$D_1$}
        & ground truth
            & -  &15.813 &-  &- &15.813 &- &- &15.813 &-
            &- &15.813 &- &- &15.813 &-\\
        & clean  
            &5.171 &24.024 &0.604 & 4.806 &23.525 & 0.587 
            &4.922 &21.345 &0.668 & 5.270 &20.208 &0.652 & 7.722 & 30.725 &0.466 \\
            
        \cdashline{2-17}[1pt/1pt]
        & random noise 
            &5.444 &27.747 &0.472 & 5.457 & 23.617 & 0.466 
            &5.397 &24.654 &0.489 & 5.503 & 33.267 & 0.449 
            & 9.853 & 49.675 &0.311 \\
        & AdvPoison~\cite{fowl2021adversarial} 
            & 10.474 & 57.699 & 0.322 & 9.310 & 27.765 & 0.347 
            & 8.069 & 35.221 & 0.393 & 8.594 & 50.696 & 0.402 
            & 13.001 &94.769 &0.190\\
        & SEP~\cite{chen2023selfensemble} 
            &8.367 &57.921 &0.321  & 7.208 & 26.003 & 0.315 
            &7.638 &\underline{62.634} &0.272
            &8.000 &55.917 &0.292 & 14.252 & 81.753 &0.209 \\
        & PTA~\cite{huang2021unlearnable}
            &\underline{11.193} &\underline{59.035} &0.286  
            &\underline{9.688} & 26.846 & 0.248  
            &\underline{9.763} &54.695 &0.242
            &\underline{10.039} &\underline{72.304} & 0.240
            &\underline{17.889} & 85.018 &0.144 \\
        & AttackVC~\cite{huang2021defending} 
            & 6.376 & 31.141 & 0.525  & 4.660 & 20.419 & 0.419  
            & 5.665 & 27.384 & 0.527 & 5.188 & 29.940 & 0.638 
            & 8.939 &62.572 & 0.289\\
        & AntiFake~\cite{yu2023antifake} 
            & 7.740 & 48.966 & \underline{0.254}  
            & 7.755 &\underline{42.890} & \underline{0.214}  
            & 6.748 & 58.420 & \underline{0.234}
            & 6.164 & 63.604 & \underline{0.221} 
            & 12.341 & \underline{98.410} & \underline{0.090}\\
        % \cline{2-14}
        \cdashline{2-17}[1pt/1pt]
        & \textbf{SPEC (ours)} 
            &\textbf{14.771} &\textbf{99.610} & \textbf{0.204}   
            &\textbf{10.278} &\textbf{57.693} &\textbf{-0.011} 
            &\textbf{13.826} &\textbf{94.706} &\textbf{0.172}
            &\textbf{11.566} &\textbf{93.270} & \textbf{0.178}
            &\textbf{22.093} &\textbf{102.407} &\textbf{0.081} \\
    \midrule
    
    \multirow{9}{*}{$D_2$}
        & ground truth
            &- &8.290 &- &- &8.290 &-  &- &8.290 &-
            &- &8.290 &- &- &8.290 &-  \\
        & clean 
            &5.629 &21.658 &0.658 & 5.079 & 6.660 & 0.561  
            &5.709 &9.700 &0.588  & 5.591 & 12.460 & 0.626 
            &7.702 &30.186 &0.425 \\

        \cdashline{2-17}[1pt/1pt]
        & random noise 
            &6.012 &26.330 &0.570 & 6.275 & 10.497 & 0.466 
            &6.103 &11.794 &0.469 &6.168 &16.491 &0.516 
            &9.586 &37.921 &0.343  \\
        & AdvPoison~\cite{fowl2021adversarial} 
            &10.438 &37.924 &0.398 
            &\underline{10.257} & 13.775 & \underline{0.292} 
            &9.150 &28.340 &0.359 &9.286 &52.709 &0.349 
            &14.207 &\underline{87.318} &\underline{0.072} \\
        & SEP~\cite{chen2023selfensemble} 
            &8.284 &\underline{50.569} &0.433 & 8.405 & 14.347 & 0.289  
            &8.390 &\underline{32.622} &0.322 &8.768 &46.412 &0.338 
            &14.423 &76.617 &0.118 \\
        & PTA~\cite{huang2021unlearnable}
            &\underline{11.504} &46.619 &\underline{0.365}  
            &9.470 &\underline{15.961} & 0.368 
            &\underline{11.041} &29.436 &\underline{0.249} 
            &\underline{12.040} &\underline{57.050} &\underline{0.280} &\underline{17.835} &82.882 &0.084\\
        \cdashline{2-17}[1pt/1pt]
        & \textbf{SPEC (ours)} 
            &\textbf{15.175} &\textbf{80.291} &\textbf{0.273} 
            &\textbf{12.303} &\textbf{16.967} &\textbf{0.267}  
            &\textbf{13.631} &\textbf{54.763} &\textbf{0.206} 
            &\textbf{13.387} &\textbf{72.909} &\textbf{0.243} 
            &\textbf{19.646} &\textbf{96.279} &\textbf{0.069} \\
    \bottomrule
  \end{tabular}
  \begin{tablenotes}
      \item $D_1$ and $D_2$ represent LibriTTS and CMU ARCTIC datasets, respectively.
  \end{tablenotes}
  \end{threeparttable}
  }
  \label{tab:1}
\end{table*}

\section{Experiments and Analyses}\label{section_experiments}
For a comprehensive conclusion, we evaluate the effectiveness, transferability, and audibility of SafeSpeech.
The effectiveness and transferability of SafeSpeech have been demonstrated against speech synthesis based on fine-tuning in Section \ref{section_exp_finetuning} and zero-shot in Section \ref{section_exp_zero-shot}, respectively. Additionally, we assess the naturalness and human perception of protected audio (in Section \ref{section_exp_finetuning} and \ref{section_exp_perception}) and conduct a subjective evaluation to investigate the deceptive effects of synthesized speech on humans. Finally, we perform ablation studies focusing on the method components and the hyperparameters.

\subsection{Effectiveness and Transferability}\label{section_exp_finetuning}
The assessment of SafeSpeech's effectiveness encompasses three different stages: the perturbation generation on the surrogate model, training on safeguarded audio samples by different methods, and evaluation of the synthetic performance.

\noindent\textbf{Perturbation Generation}. We select BERT-VITS2 as the surrogate model to generate perturbation due to its superior performance in high-quality speech synthesis and fine-tuning capability. We utilize the surrogate model to generate perturbation via SafeSpeech and evaluate its transferability on other SOTA TTS models. For a comprehensive evaluation, we select six baseline methods, including random noise and the specifically generated perturbation.

\noindent\textbf{Training on the Safeguarded Dataset}. After acquiring specific noises from the surrogate model, we apply the perturbation to the original audio, creating the protected dataset. Fine-tuning on unprotected audio samples results in a plausible speaker synthesizer. We compare this scenario with SafeSpeech-protected audio and verify their unlearnability across StyleTTS2, MB-iSTFT-VITS, VITS, and GlowTTS without \textit{a priori} knowledge of the model structures.

\noindent\textbf{Speech Synthesis and Evaluation Results} After training, we assess the synthesizer's performance on the test set. By inputting a speaker ID and text, the synthesizer produces realistic deepfake audio. For each test sample, we supply the generator with speaker ID and text, yielding synthesized speech. We evaluate it by measuring MCD and SIM between real and synthesized speech and using WER to measure speech clarity.

\begin{table}[t]
    \centering
    \caption{Objective evaluation of the similarity and naturalness between protected and original real audio samples.}
    \setlength\tabcolsep{2pt}
    \setlength{\extrarowheight}{5pt}
    \resizebox{0.85\linewidth}{!}{
    \begin{tabular}{c|ccccc:c}
        \hline
        & AdvPoison 
        & SEP & PTA & AntiFake & AttackVC & \textbf{SafeSpeech}\\
        \hline
        \textbf{Similarity($\uparrow$)}
            & 0.715 & 0.703 & 0.776 & 0.719 
            & \textbf{0.974} & \underline{0.859} \\
        \hline
        \textbf{Naturalness($\uparrow$)} 
            & \underline{3.343} & 2.571 & 2.515 & 2.824 
            & \textbf{4.289} & 3.021\\
        \hline
    \end{tabular}
    }
    \label{tab_sim_mos}
\end{table}

\noindent\textbf{Effectiveness and Transferability Analyses.~} Table \ref{tab:1} shows the experimental results. Our proposed SPEC has achieved excellent protection performance on both single-speaker and multi-speaker datasets in terms of timbre (SIM) and speech intelligibility (MCD and WER). 
On the LibriTTS dataset, the SPEC effectively safeguards speeches from being learned with a significant increase in WER from 24.024\% of the clean dataset to 99.610\%.
High WER values represent low speech quality (SA2). And SIM is the lowest at 0.204 which satisfies SA1. Moreover, the results show a broad range of transferability across TTS models with distinct structures. 
Compared to the PAP methods for data protection during the training stage and voice protection techniques for the inference stage and speaker similarity, the SPEC approach demonstrates superior performance in preventing the usage and high similarity of the synthesized audio with outstanding transferability. The reasons encompass two aspects. (1) The {\it \uline{pivotal objective optimization}} is a universal objective, ensuring better effectiveness and transferability. (2) The \uline{\textit{SPEC}} technique can effectively conceal the speaker's information, thereby successfully preventing the model from learning audio samples.

\noindent\textbf{Perception Analyses.} In the design of SafeSpeech, we optimize the perception of the perturbation in the time and frequency domain (Section \ref{section_method_perception}). It is crucial that the perturbation cannot affect the normal use of the audio or alter the timbre. Therefore, we objectively evaluate the best speaker similarity (the SIM metric) and naturalness between the protected and clean audio. Table \ref{tab_sim_mos} illustrates the results of our experiments, which show that compared to better-performing baselines, \textit{e.g}, PTA and AntiFake, SafeSpeech achieves a similarity score of 0.859, indicating almost no alteration in the timbre, and a naturalness score of 3.021. Moreover, we balance the effectiveness and perception by sampling $\alpha$ in Section \ref{section_exp_ablation}.

This experiment confirms that SafeSpeech is effective and transferable with minimal alteration of the original audio.

\begin{table}
    \centering
    \begin{minipage}{0.24\textwidth}
        \centering
        \caption{{The subjective evaluation of the ground truth (GT) and synthesized speech.}}
        \label{tab_perception_effect}
        \setlength\tabcolsep{2pt}
        \resizebox{0.9\textwidth}{!}{
        \begin{tabular}{c|c}
            \hline
            & {\bf MOS}($\downarrow$) \\
            \hline
            GT & 4.756 $\pm$ 0.103 \\
            \hline
            clean & 4.677 $\pm$ 0.114 \\
            \hline
            PTA & 2.008 $\pm$ 0.191  \\
            \hline
            {\bf SPEC (ours)} & {\bf 1.070 $\pm$ 0.161}  \\
            \hline
        \end{tabular}
        }
    \end{minipage}
    \quad
    \begin{minipage}{0.20\textwidth}
        \centering
        \caption{{Human perceptual evaluation of the similarity and naturalness between protected and original real audio samples.}}
        \label{tab_perception_sim_na}
        \setlength\tabcolsep{2pt}
        \resizebox{0.9\textwidth}{!}{
        \begin{tabular}{c|c}
            \hline
            {\bf Similarity} & {\bf Naturalness} \\
            \hline
            98.333\% & 3.190 $\pm $ 0.189\\
            \hline
        \end{tabular}
        }
    \end{minipage}
\end{table}

\subsection{User Study}\label{section_exp_perception}
In the real world, deepfake speech usually needs to deceive human victims. Therefore, in this section, we explore the human perception of protected samples and synthetic speech.

\noindent\textbf{Preliminary Work.} The Human Ethics Research Committee affiliated with the primary author determined that this study was exempt from further human subject review. We created the anonymous questionnaire and recruited participants through the Credamo platform.

\noindent\textbf{Participants.} We recruited 80 participants (after filtering in Appendix \ref{section_append_user}), all of whom were between 18 and 40 years old and had proficient English skills. Before participating, we obtained their consent and provided an average compensation of \$0.30 per participant. Their average spent time is 241.075 seconds, providing reliable subjective results.

\noindent\textbf{Study Setting.} To throughout evaluate, we have designed three parts in each questionnaire with 23 questions to explore the synthesis quality, speaker similarity between protected and original audio, and the naturalness of protected audio.

\noindent\textbf{Part 1: Synthesis Quality.} We selected three ground truth and synthesized audio from the clean, better-performing baseline (PTA), and SafeSpeech-protected datasets (twelve samples in total) to validate the subjective quality. The participants rated the audio quality on a scale from 0 to 5. Table \ref{tab_perception_effect} shows the results. Clean synthesized samples have the MOS value of 4.677$\pm$0.114, indicating good audio perceptual quality and the potential to deceive participants. In contrast, the MOS for the audio synthesized from the SafeSpeech-protected dataset is much lower at 1.070$\pm$0.161, reflecting the poor audio quality and inability to deceive participants effectively.

\noindent\textbf{Part 2: Speaker Similarity.} The experiments in Part 1 demonstrate the perceptual effectiveness of SafeSpeech and we also consider if the protected audio retains a similar timbre to the original. We select three pairs of protected audio and original audio with one pair from different speakers for the experiment. Table \ref{tab_perception_sim_na} shows that 98.333\% of participants believe the protected audio came from the same speaker as the original, indicating minimal alteration to the original speaker's timbre.

\noindent\textbf{Part 3: Naturalness of Protected Audio.} We aim to degrade the detectability of human perception. In this part, we ask participants to rate the naturalness of the protected audio from 0 to 5~\cite{yu2023antifake}. Table \ref{tab_perception_sim_na} shows that the naturalness score for the protected audio is 3.190$\pm$0.189. Generally, a score above 3 is considered to indicate relatively high quality~\cite{yu2023antifake}. Therefore, most participants find the protected audio is natural or the embedded perturbation is acceptable.

\subsection{Defense against Zero-Shot Voice Cloning}\label{section_exp_zero-shot}
When adversaries derive the audio samples of the target speaker, they may apply them to fine-tune a synthesizer or conduct zero-shot cloning with a limited number of samples. Zero-shot voice cloning requires fewer computational resources than fine-tuning but degrades the synthetic result. The experiments in Section \ref{section_exp_finetuning} demonstrate SafeSpeech's effectiveness at training time. However, we also aim SafeSpeech can still perform well against zero-shot voice cloning. In this section, we utilize five advanced SOTA TTS models, \textit{i.e.}, TorToise-TTS~\cite{betker2023better}, XTTS~\cite{casanova2024xtts}, OpenVoice~\cite{qin2023openvoice}, FishSpeech~\cite{liao2024fish}, and F5-TTS~\cite{chen2024f5}, with outstanding zero-shot capability to evaluate the synthesis on both clean and protected samples.

\begin{figure}[t]
\centerline{\includegraphics[width=0.3\textwidth]{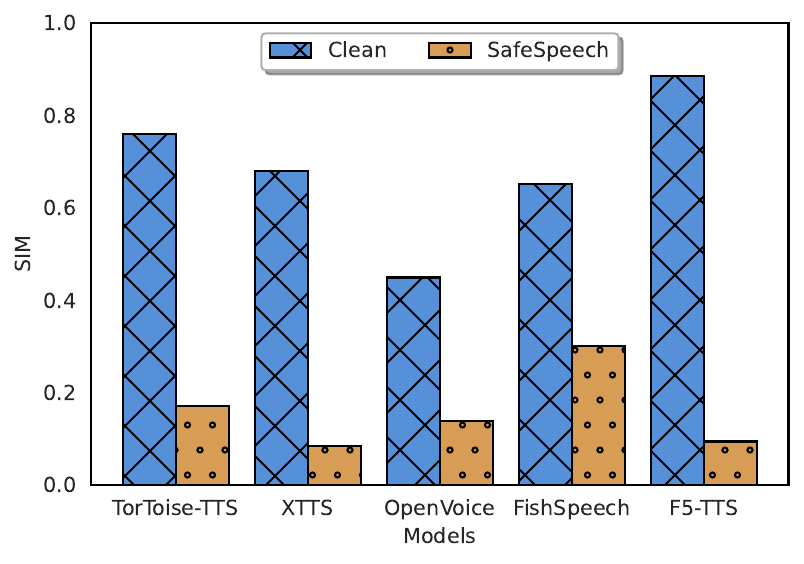}}
\caption{Speaker similarity of synthesized samples after zero-shot voice cloning on the clean and protected audio.}
\label{fig_zeroshot}
\end{figure}

\noindent\textbf{Experiments and Results.} Figure \ref{fig_zeroshot} presents the speaker similarity, \textit{i.e.}, the SIM metric, between the clean and SafeSpeech-protected synthesized speech to real audio. Surprisingly, although the generation process of perturbations does not depend on these zero-shot models, and the perturbation is specifically considered for the training phase, SafeSpeech can still protect our speech during the inference procedure on the unseen and advanced models. On the F5-TTS, the SIM value drops from 0.885 for clean samples to 0.094 for protected samples. Even the FishSpeech, which exhibits the best performance against noise, achieves only a SIM of 0.301 on protected samples. This indicates that SafeSpeech remains effective in defending against zero-shot voice cloning.

\noindent\textbf{Analyses.} This effectiveness is due to our proposed SPEC method in Section \ref{section_method_spec}, which leverages the surrogate model to guide the synthesized speech more noise-like via the KL divergence, thereby concealing the original speaker's information. Consequently, this approach ensures protection during both the fine-tuning and zero-shot stages.

\subsection{Ablation Study}\label{section_exp_ablation}
In Section \ref{section_method_spec}, we delve into the rationale behind objective selection. Building upon this foundation, this section presents a comprehensive comparison of the original and pivotal objective optimization methods, evaluating their effectiveness and runtime performance in perturbation optimization. We achieve a balance of effectiveness and perception of perturbations by sampling $\alpha$ in Eq. (\ref{eq_protection}).  Meanwhile, our proposed SPEC is a multi-task learning problem, so the setting of $\beta$ is an issue of interest. We conduct the ablation study on the LibriTTS dataset and the BERT-VITS2 model.

\begin{table}[t]
  \centering
  \caption{The difference between the vanilla and our proposed pivotal function as optimization function.}
  \resizebox{0.9\linewidth}{!}{
  \begin{tabular}{cccccccc}
    \toprule
    \multirow{2}{*}{Method} &\multirow{2}{*}{\# Params}
    & \multicolumn{2}{c}{Vanilla} & \multicolumn{2}{c}{\textbf{Pivotal (ours)}} \\
    \cmidrule(r){3-4}\cmidrule(lr){5-6}
    & & MCD($\uparrow$)   & WER($\uparrow$) 
    & MCD($\uparrow$)   & WER($\uparrow$)\\
    \midrule
    BERT-VITS2  &104.64 M &10.316 &72.435  
        &\textbf{10.722} &\textbf{81.517} \\
    MB-iSTFT-VITS & 80.78 M &9.074 &64.421 
        &\textbf{9.945} &\textbf{73.756} \\
    VITS & 82.42 M &9.773 &64.614 
        &\textbf{10.419} &\textbf{87.846}\\
    GlowTTS& 32.03 M  &17.113 & 95.632 
        & \textbf{18.607} &\textbf{104.887}\\
    \bottomrule
    \end{tabular}
    }
    \label{table_vanilla}
\end{table}

\noindent\textbf{Efficiency and Effectiveness of Pivotal Objective}. In Section \ref{section_method_spec}, we have introduced the problem formulation and illustrated the comparison between vanilla and pivotal unlearnable examples. To address a multi-task optimization problem, we simplify it into a single-task optimization problem and illustrate the principles of the pivotal function selection strategy. Taking the BERT-VITS2 model as an example, through the analysis of the convergence speed, we devise a universal function that is most relevant to the audio content as part of our target optimization function. This section compares using the $\mathcal{L}_{mel}$ as the primary objective of optimizing the entire generator functions in terms of efficiency and time cost.

Table \ref{table_vanilla} shows our experimental results. Among them, on BERT-VITS2 we find that the WER increases from 72.435\% to 81.656\%, which means less clear speech expression and noisy background. We find that all have certain unlearnability across three models. While achieving better protection, choosing the pivotal function to optimize can greatly shorten the noise optimization runtime. In our experiments, the time for optimizing one identical sample to generate vanilla perturbation is 10.3 seconds, whereas the time required for pivotal optimized perturbation is 4.0 seconds, resulting in a nearly 61.2\% reduction which can be employed for real-world application. The improved efficiency is due to the pivotal function optimization can avoid useless calculations.

\begin{figure}[t]
    \centering
    \begin{subfigure}{0.20\textwidth}
        \centering
        \includegraphics[width=\textwidth]{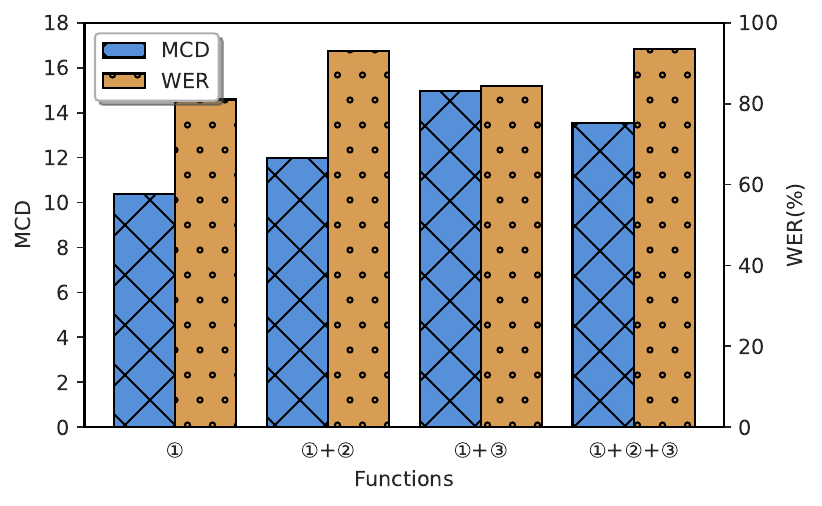} 
        \caption{Component comparison.}
        \label{fig_ablation_a}
    \end{subfigure}
    \begin{subfigure}{0.20\textwidth}
        \centering
        \includegraphics[width=\textwidth]{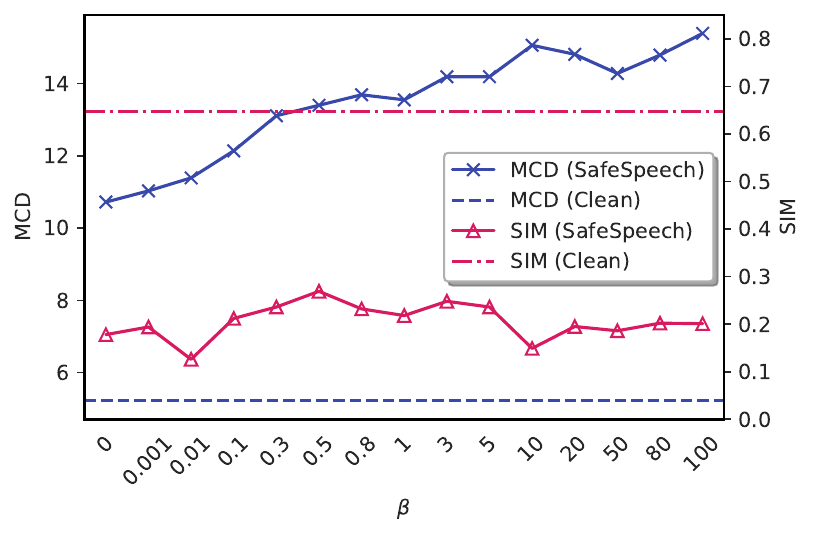}
        \caption{Parameter $\beta$ comparison.}
        \label{fig_ablation_b}
    \end{subfigure}
    \begin{subfigure}{0.45\textwidth}
        \centering
        \includegraphics[width=\textwidth]{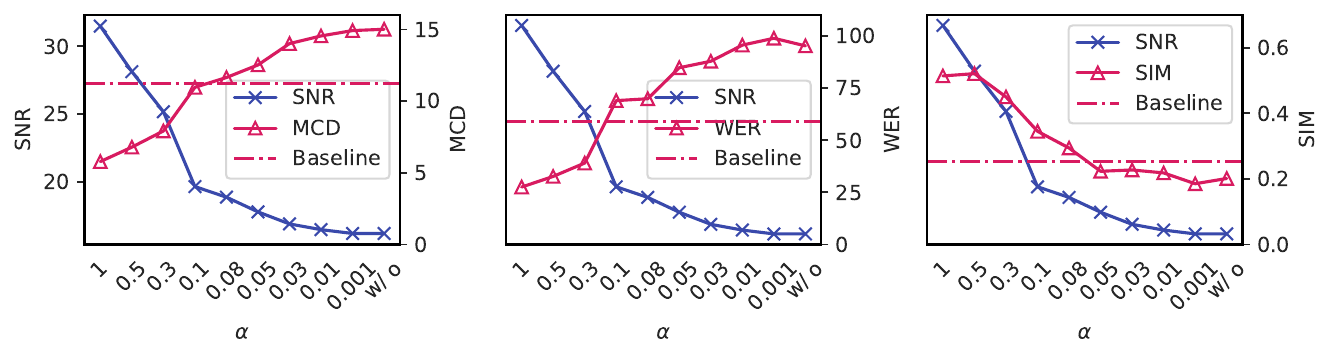} 
        \caption{Comparison of perception and effect across different $\alpha$.}
        \label{fig_ablation_alpha}
    \end{subfigure}
    \caption{Ablation study about components analyses and hyperparameter settings on different evaluation metrics.}
\end{figure}

\noindent\textbf{Components Analyses.} The objective function of SPEC is described as Eq. (\ref{eq_SPEC}) which in detail can be divided into three parts: \circled{1} the pivotal function; \circled{2} the KL divergence using Gaussian noise to lead the noise-like output; \circled{3} the $\ell_1$ norm between random noise and synthetic audio. To investigate how each function affects the protective effect, we carry out the ablation study on the LibriTTS dataset, considering the combination of functions: only \circled{1}, \circled{1}+\circled{2}, \circled{1}+\circled{3} and \circled{1}+\circled{2}+\circled{3}, respectively.  The results in Figure \ref{fig_ablation_a} demonstrate the different effectiveness of the four functions. Among them, the introduction of noise-leading methods in \circled{2} and \circled{3} both yield better results than using the pivotal loss function \circled{1} alone. Additionally, we find that the combination of the three functions, \circled{1}+\circled{2}+\circled{3}, performs best on WER and the SIM value is only 0.21, which is lower than the speaker similarity threshold, indicating outstanding protection of the speaker's timbre when combining the three functions as Eq. (\ref{eq_SPEC}).

\noindent\textbf{Balance of Strength and Perception.} In Section \ref{section_method_perception}, we introduce a perceptual loss based on two evaluation metrics for perturbation, \textit{i.e.}, STFT and STOI, to enhance the imperceptibility in the time and frequency domain. Eq. (\ref{eq_protection}) shows the optimization objective of SafeSpeech, while the value of $\alpha$ influences the effectiveness of protection and imperceptibility of the perturbation: a larger $\alpha$ results in better auditory quality but weaker protective effect. We explore the balance by sampling $\alpha$ from 0.001 to 1 with the SNR as the perceptual metric. Figure \ref{fig_ablation_alpha} illustrates the results. We can find that when $\alpha$ is set to 0.05, the effectiveness metrics are higher than ``Baseline'' (here we select PTA), \textit{i.e.}, resulting in an MCD of 12.516, a WER of 84.709\%, and a SIM of 0.223. The perceptual metric increases from an initial 16.021 to 17.791, which also surpasses the ``Baseline'' score of 16.578.

\noindent\textbf{Hyperparameter Study.} Our proposed objective Eq. (\ref{eq_SPEC}) shows a multi-task optimization problem, in which the value of the weight coefficient $\beta$ has a certain impact on the performance of noise optimization. In this experiment, we study how different values of $\beta$ impact the unlearnability of protected audio. We carry out experiments on a single speaker from LibriTTS and fine-tune the utterances for 100 iterations. We establish a range for $\beta$ from 0 to 100, with various intervals, resulting in a total of fifteen distinct $\beta$ values. Figure \ref{fig_ablation_b} shows the results of different $\beta$ on the MCD and SIM metrics.  We observe that the impact of unlearnability remains consistent across different values of $\beta$, enabling us to achieve satisfactory results regardless of its specific setting. We find that when $\beta$ is set to 0.01 or 10, the timbre is protected achieving well, while the value of 10 also poisoning the dataset with a higher MCD of 15.060 compared to 11.356. On this basis, we have chosen $\beta$ to be 10, which yields WER and SIM values of 95.266\% and 0.149, respectively. When compared to the training on clean samples, the MCD and SIM values stand at 5.217 and 0.648, respectively, thus demonstrating a remarkable protection performance. The insensitivity of the protective effect to the hyperparameter indicates the stability.

\section{Robustness against Adaptive Attackers}\label{section_robustness}
SafeSpeech possesses high robustness against strong adaptive adversaries. In this section, we consider and conclude three levels of adversaries: (1) \underline{\it Data Level}. Data-level technologies include perturbation removal in Section \ref{section_robustness_removal}, advanced data augmentation in Section \ref{section_robust_aug}, and optimization-based speech recovery in Section \ref{section_robust_speech_recover}. (2) \underline{\it Model Level}. Adversaries may employ model recovery in Section \ref{section_robust_recovery}, robust training in Section \ref{section_robust_advtrain}, and fine-tuning with clean data in Section \ref{section_robustness_finetunig}. (3) {\it \uline{Real-world Level}}. Protection in the physical world is also a requirement at a higher level. We will evaluate the robustness of SafeSpeech in the physical world in Section \ref{section_robust_real-world}, as well as the performance and time overhead in real-time scenarios.

\subsection{Data-Level Robustness}
\subsubsection{Advanced Perturbation Removal}\label{section_robustness_removal}
The embedded perturbation for protecting audio by SafeSpeech may be detected by adversaries, and they can remove this noise to improve the synthetic performance. In this section, we consider the traditional denoising technique, spectral gating (SG), as well as the current mainstream and advanced denoising model based on deep learning, DEMUCS~\cite{rouard2022hybrid}.

\noindent\textbf{Traditional Denoise.} SG aims to remove the relatively low value in the time domain of the speech. After training with SG denoising on audio protected by SafeSpeech, the results show a WER of \textbf{69.321\%}, and a SIM of \textbf{0.233}, indicating that SafeSpeech remains relatively effective under this condition.

\noindent\textbf{Advanced Denoise.} DEMUCS can effectively eliminate perturbations and obtain approximately clean samples without a noisy background. We train the model using DEMUCS-denoised samples, resulting in a WER of \textbf{57.329\%} and a SIM of \textbf{0.284}. This indicates that the synthesized speech is dissimilar and of low quality compared to the original samples.

\noindent\textbf{Reason and Analyses.} SafeSpeech is robust against traditional and advanced perturbation removal techniques. This is because SafeSpeech embeds imperceptible perturbations, while denoising can remove noise along with some original speaker information, \textit{e.g.}, timbre, and phoneme features. 

\begin{table}[t]
    \centering
    \caption{The robustness quantization via data augmentation and defensive methods. The \underline{underline} values indicate the most significant decreases in protection compared to training without data augmentation, \textit{i.e.}, SPEC (``w/ o'' in the Table).}
    \setlength\tabcolsep{2pt}
    \resizebox{\linewidth}{!}{
    \begin{tabular}{ccccc ccccc cccccc}
      \toprule
      \multirow{2}[2]{*}{Metric}
      & \multirow{2}[2]{*}{w/ o} & \multicolumn{4}{c}{Defense-based~\cite{hussain2021waveguard}}
      & \multicolumn{4}{c}{Transformation-based} 
      & \multicolumn{1}{c}{Diffusion-based}\\
      \cmidrule(l){3-6}\cmidrule(lr){7-10}\cmidrule(r){11-11}
      & & RS & Mel & QD & FL 
      & Speed & Mask & LPF & MP3 & AP~\cite{wu2023defending}\\
      \midrule
      MCD($\uparrow$) 
        & 14.771 & 14.368 & 14.679 & 14.486 & 13.222
        & \underline{11.455} & 14.983 & 13.991 & 15.097 & 14.216\\ 
      WER($\uparrow$) 
        & 99.610 & 96.781 & 99.149 & 91.363 & 97.306
        & 97.444 & 100.899 & 98.082 & 93.545 & \underline{85.711}\\
      SIM($\downarrow$)
        & 0.204 & 0.168 & 0.179 & 0.238 & 0.227
        & 0.106 & 0.247 & 0.252 & \underline{0.261} & 0.227 \\
      \bottomrule
    \end{tabular}
    }
    \label{tab_trans}
  \end{table}

\subsubsection{Data Augmentation}\label{section_robust_aug}
In the real world, the attackers may adopt various data augmentation methods to destroy the specific perturbation to improve the model performance, so we hope that users' uploaded audio protected by SafeSpeech can retain the consistency of unlearnability when facing real-world speech synthesis attacks with different data augmentation methods. For a comprehensive evaluation, following \cite{yu2023antifake, chen2024songbsab}, we consider three categories of effective data augmentation techniques: defense-based, transformation-based, and diffusion-based.

\begin{itemize}[leftmargin=1em, topsep=0pt, itemsep=0pt, parsep=0pt]
    \item \textbf{Defense-based Techniques}: These include down-sampling and up-sampling (RS), mel-spectrogram extraction and inversion (Mel), quantization-dequantization (QD), and frequency filtering (FL), which are designed to effectively prevent adversarial audio examples from WaveGuard~\cite{hussain2021waveguard}.
    \item \textbf{Transformation-based Techniques}: This category encompasses speed adjustment, time masking, low-pass filtering (LPF), and MP3 compression, which are commonly applied in real-world audio processing.
    \item \textbf{Diffusion-based Techniques}: AudioPure~\cite{wu2023defending}. AudioPure (AP) aims to disrupt the perturbed audio by diffusion model, which represents the SOTA audio defense technique.    
\end{itemize}

From the results in Table \ref{tab_trans}, we can observe that these data augmentation techniques diminish the protective effect to a certain extent. For instance, the WER decreases to 85.711\% by AudioPure, and the SIM metric drops to 0.261 by MP3 Compression. However, compared to clean samples or those with random noise, the protection remains effective. This indicates that SafeSpeech demonstrates robustness against audio data augmentation techniques. The reason for this robustness is that while data augmentation can disrupt the structures of embedded perturbations, these transformations can also degrade speech quality (\textit{e.g.}, MP3 Compression, mel extraction, and speed adjustment). Consequently, lower-quality speech samples as input will result in the degradation of both the quality and similarity of the synthesized speeches.

\subsubsection{Optimization-based Speech Recovery}\label{section_robust_speech_recover}
The adversaries may attempt to recover the original speech from the perturbed state in a ``reverse'' direction as SafeSpeech, with the critical challenge of determining the ``reverse'' optimization direction. However, the optimization direction remains unknown due to the lack of clean audio from the original speaker. Further analysis reveals that SafeSpeech aims to conceal the speaker's privacy by embedding perturbations. Therefore, the adversaries may leverage two types of feedback to determine the optimization direction~\cite{chen2024songbsab, yu2023antifake}: (1) The naturalness score, aiming to restore it to an unperturbed state; (2) By increasing adversaries' capabilities, they can query a speaker recognition system enrolled with the target speaker, to recover the hidden characteristics.

Following~\cite{chen2021real, ilyas2018black}, we employ a black-box optimization method, Natural Evolution Strategies (NES), with a total of 50000 queries to optimize a random sample from the LibriTTS dataset. Then, the reversed sample is cloned by FishSpeech, resulting in a SIM value of 0.252. Compared to the initial perturbed sample of a SIM of 0.168, this optimization-based method can improve the synthetic performance but is still far from the original sample's SIM value of 0.561. The reasons lie in two aspects. Firstly, adding perturbation during this optimization degrades the speech quality. Secondly, the ``reverse'' direction is estimated and not inaccurate.

\subsection{Model-Level Robustness}
\subsubsection{Model Recovery}\label{section_robust_recovery}
In this experiment, we verify the necessity of fine-tuning for the target speaker and the feasibility of the adversary's model recovery technique by effectiveness comparison.

Before performing speech synthesis, the adversary already possesses a model pre-trained on a large-scale dataset, and subsequently acquires a protected dataset for fine-tuning, with the hope of cloning the target speaker. However, after training, they are unable to clone high-quality audio and realize that the model may have been perturbed. Consequently, they can {\it \uline{recover from using the original model}} for voice generation, so this experiment considers the model scenario and parameter restoration. We fine-tune the pre-trained model, BERT-VITS2, and validate the untrained synthetic voice against the LibriTTS speaker. The results are an \textbf{MCD of 14.911, a WER of 100\%, and a SIM of 0.049}, indicating that without fine-tuning, the generated audio is not similar compared to fine-tuning of MCD at 5.171, WER at 24.024\%, and SIM at 0.604. Therefore, direct synthesis without fine-tuning cannot achieve an effective synthetic result for the target victim.

We also find that the WER value is relatively high, which is because the released BERT-VITS2 model is trained on a large-scale dataset without setting the exact speaker number, while our focus is on fine-tuning for a single speaker. Therefore, it cannot completely load the weight files; only by fine-tuning the pre-trained model can we generate audible audio.

\subsubsection{Advanced Robust Training}\label{section_robust_advtrain}
Previous PAP methods show the vulnerability against robust training techniques~\cite{huang2021unlearnable, fu2022robust}. 
In this section, we employ adversarial training to illustrate the robustness of SafeSpeech.

\noindent\textbf{Adversarial Training.~} Adversarial training aims to generate adversarial perturbations by maximizing errors and incorporating them into training samples to enhance the model's robustness and performance. After the adversary obtains the protected samples, they can maximize the objective function in Eq. (\ref{eq_protection}) to generate adversarial perturbation that can mitigate SafeSpeech. Meanwhile, previous perturbative defense mechanisms that generate perturbations based on $\ell_p$ norm are vulnerable to attacks if the adversarial perturbation radius $\rho_a$ exceeds the defensive data perturbation radius $\rho_u$. In such cases, the effectiveness of data protection significantly degrades, making non-robust data protection methods easy to compromise. In this section, for a comprehensive evaluation of the robustness of data protection, we set $\rho_u$ to $8/255$ and $4/255$ respectively, while the $\rho_a$ is varied across a range of values: $0, 2/255, 4/255, 8/255, 10/255, 12/255, 16/255$, to conduct a thorough assessment of the adversarial training.

\begin{table}[t]
  \centering
  \caption{The model performance across different defensive perturbation radius $\rho_u$ and adversarial perturbation radius $\rho_a$.}
  \resizebox{0.9\linewidth}{!}{
  \begin{tabular}{cccccccc}
    \toprule
    \multirow{2}{*}{$\rho_a$}
    & \multicolumn{3}{c}{$\rho_u = 8/255$} & \multicolumn{3}{c}{$\rho_u = 4/255$} \\
    \cmidrule(r){2-4}\cmidrule(lr){5-7}
    & MCD($\uparrow$)   & WER(\%)($\uparrow$)  & SIM($\downarrow$)
    & MCD($\uparrow$)   & WER(\%)($\uparrow$)  & SIM($\downarrow$)\\
    \midrule
    $0$      &14.771 &99.610 &0.204 &10.921 &76.110 &0.302 \\
    $2/255$  &14.592 &92.891 &0.214 &8.538 &55.581 &0.292\\
    $4/255$  &12.361 &99.079 &0.147 &\textbf{6.554} &\textbf{55.127} &\textbf{0.315}\\
    $8/255$  &11.323 &\textbf{82.504} &0.188 &7.946 &68.467 &0.246\\
    $10/255$ &\textbf{11.060} &94.608 &0.198 &8.621 &80.703 &0.225\\
    $12/255$ &11.771 &107.829 &\textbf{0.220} &9.086 &84.630 &0.243 \\
    $16/255$ &12.262 &113.238 &0.193 &10.304 &94.940 &0.237\\
    \bottomrule
  \end{tabular}
  }
  \label{tab_adv}
\end{table}

From Table \ref{tab_adv}, we can find that when $\rho_a$ is greater than or equal to $\rho_u$, the data protection effect can be reduced. For example, when $\rho_a$ and $\rho_u$ are both $8/255$, MCD and WER decreased from 14.716 and 97.090\% to 11.323 and 82.504\% respectively. However, the attack result of model performance improvement is still not obvious, far higher than the threshold. This proves that in the face of adversarial training, whether it is less than, equal to, or greater than the $\ell_p$ radius, the attacker still cannot maliciously clone the protected data.

\subsubsection{Fine-tuning with Clean Data}\label{section_robustness_finetunig}
In this section, we explore if SafeSpeech can be circumvented by clean-data fine-tuning and analyze its difference from data poisoning. Adversaries on the Internet may obtain both clean and SafeSpeech-protected audio samples and employ the mixed data for training to boost efficiency. If the model performs poorly on a certain speaker after training, it may suggest the speaker is protected. Then, they may use clean-data fine-tuning to reduce the speaker's impact on the model and try to recover the model performance. 

\noindent\textbf{Training with mixed data.} We randomly select five speakers from the LibriTTS dataset ~\cite{zen2019libritts}, where speaker $i$ is protected, and the remaining four are clean speakers, forming a mixed dataset for training on BERT-VITS2. The protected samples account for approximately 14.6\% (108 out of 738) of this dataset. After training, the results present that cloning speaker $i$ yielded a WER of 108.336\% and a SIM of 0.221 while testing with clean speakers resulted in a WER of 16.506\% and a SIM of 0.686. This indicates that during joint mixed training, the protected samples do not interfere with the clean samples, nor do the clean samples mitigate the protected samples. Therefore, SafeSpeech only affects the audio that needs protection, which is distinctly different from data poisoning. Data poisoning aims to degrade model performance, meaning perturbed audio would affect the clean samples.

\noindent\textbf{Fine-tuning with clean data.} The aforementioned scenario illustrates that audio protected by SafeSpeech is not mitigated by clean samples trained alongside it. Therefore, adversaries may obtain additional clean samples for further fine-tuning to mitigate the perturbed samples and recover the model's performance. We randomly select five new speakers from the LibriTTS dataset to form a clean sample dataset for fine-tuning, with a total of 616 training samples. After fine-tuning, we test models on speaker $i$ and result in a WER of 37.050\% and a SIM of 0.126. It can be observed that fine-tuning with clean data leads to some improvement in clarity, but the similarity is highly low. This is because fine-tuning with clean data is conducted after training speaker $i$ which means fine-tuning overwrites the original speaker's timbre and replaces it with the characteristics of the fine-tuned samples, \textit{i.e.}, previous speakers have not been ``learned'' to achieve effective cloning.

\subsection{Real-World Robustness}\label{section_robust_real-world}
In real-world scenarios, \textit{e.g.}, personal on-site presentations or online live broadcasts, (near) real-time and effective protection is required. In this section, we test the robustness and time overhead under real-time requirements in the real world. We utilize a lightweight TTS model, MB-iSTFT-VITS~\cite{kawamura2023lightweight}, as the surrogate model for perturbation generation.

\noindent\textbf{Universality.} For a single speaker, we generate perturbation from an audio sample and apply it to pad or truncate other samples.  After fine-tuning with MB-iSTFT-VITS, the WER is 72.472\%, and SIM is 0.197, indicating SafeSpeech can protect using just one segment of the target speaker's audio.

\noindent\textbf{Real-World Protection.} We invite a volunteer to read LibriTTS texts in a quiet room for 10 minutes for each test. The initial volume of the room is 22 dBA. We deploy SafeSpeech on a GPU-equipped device as the back end. The front end is a Lenovo laptop with an Intel(R) microphone for recording and a Lenovo BMS09 speaker for playing noise as the defender. The whole process is: When the microphone captures about 5 seconds of audio containing the target speaker, the front end sends the recorded audio to SafeSpeech's back end ({\it \uline{speaker $\to$ microphone $\to$ SafeSpeech}}). SafeSpeech then generates perturbations from it and continuously sends perturbations to the front-end speaker for playback ({\it \uline{SafeSpeech $\to$ speaker}}), thus achieving real-time protection in the real world. Given that recording in real-world scenarios may suffer quality degradation, we apply SG denoise and ``loud-norm''~\cite{vickers2010loudness} to the audio received by SafeSpeech to enhance vocal clarity. Meanwhile, an adversary records the live sound from a distance of 50 cm using a mobile device, Android VIVO.

\begin{figure}[t]
    \centering
    \begin{subfigure}{0.155\textwidth}
        \centering
        \includegraphics[width=\textwidth]{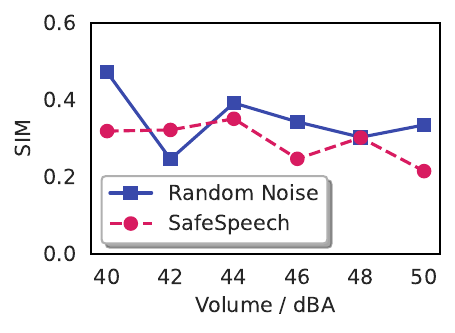} 
        \caption{FishSpeech.}
        \label{fig_real-world_a}
    \end{subfigure}
    \begin{subfigure}{0.155\textwidth}
        \centering
        \includegraphics[width=\textwidth]{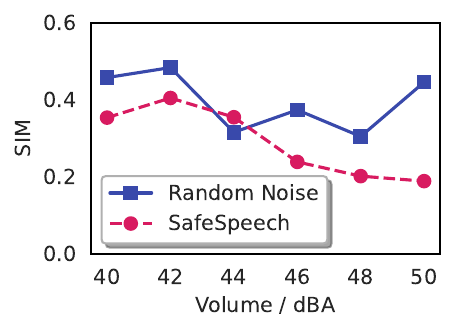}
        \caption{XTTS.}
        \label{fig_real-world_b}
    \end{subfigure}
    \begin{subfigure}{0.155\textwidth}
        \centering
        \includegraphics[width=\textwidth]{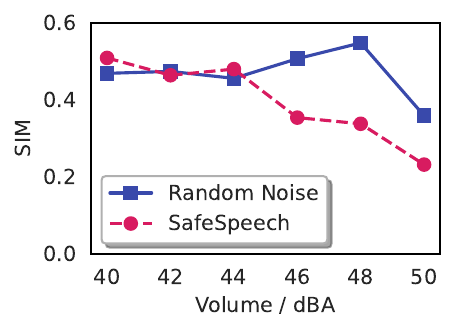}
        \caption{F5-TTS.}
        \label{fig_real-world_c}
    \end{subfigure}
    \caption{The results of synthetic speaker similarity in the physical world across three TTS models and volumes.}
    \label{fig_real-world}
    \vspace{-2mm}
\end{figure}

\noindent\textbf{Results and Analyses.} Referring to VSMask~\cite{wang2023vsmask}, we evaluate the protection effectiveness of perturbations at volumes ranging from 40 dBA to 50 dBA, with random noise added as a reference. We test the volumes for ten seconds and calculate the averages. We conduct voice cloning tests on three models, \textit{i.e.}, FishSpeech~\cite{liao2024fish}, XTTS~\cite{casanova2024xtts}, and F5-TTS~\cite{chen2024f5}. Figure \ref{fig_real-world} illustrates the performance in the real world. It shows that at 40 dBA, SafeSpeech can resist cloning attempts of FishSpeech and XTTS, whereas, at 50 dBA, these three models cannot effectively clone the speaker with a SIM of 0.215 in FishSpeech. Moreover, we find that SafeSpeech outperforms random noise. This experiment demonstrates SafeSpeech's robustness in the real world. The reason for robustness lies that real-world recording is equivalent to a transformation, and SafeSpeech can resist data augmentation (Section \ref{section_robust_aug}).

\noindent\textbf{Time Overhead.} We build SafeSpeech on an NVIDIA A800 GPU device and average results over ten runs for reliability. The whole process, from getting the initial audio to complete playback, takes 13.898 seconds. It takes 10.606 seconds to generate perturbation for the speaker and 0.369 seconds to transmit the noise via various devices and networks.
Compared to VSMask~\cite{wang2023vsmask}, a real-time voice defender that takes about 300 seconds to predict a speaker, SafeSpeech can protect on average in just a 14-second lead time for continuous protection. {\it \uline{The outstanding real-time capability comes from the pivotal function of decreasing the computational time and the choice of a lightweight surrogate model.}}

\section{Discussions and Limitations}\label{section_discussion}
In this section, we discuss some unavoidable points.

\noindent\textbf{Distinction from Data Poisoning.~} We aim that the protected data cannot be learned by the TTS models. The experiment in Section \ref{section_robustness_finetunig} illustrates this point. If an adversary unauthorizedly obtains the target speaker's voice in a batch of data, this batch of data cannot be learned or cloned, and it does not affect the use of other authorized data. Data poisoning aims to degrade the model's performance. Although it protects unauthorized use to some extent, it also interferes with the use of authorized data, thus affecting the right of other data to be used, which is not our intention. Although SafeSpeech is like data poisoning, our purpose and results are quite different.

\noindent\textbf{Benifits.~} Compared to adversarial-example-based voice protection techniques~\cite{yu2023antifake, wang2023vsmask, huang2021defending}, we propose the pivotal objective optimization based on unlearnable examples that can effectively achieve training stage protection with a broader application rather than zero-shot. Compared to PAP methods~\cite{huang2021unlearnable, chen2023selfensemble, fowl2021adversarial}, we introduce the SPEC technique based on KL divergence to guide the model output towards noise audio with the actual speaker information seemly ``Consealing''.

\noindent\textbf{Further Effectiveness Enhancement.} In the future, we can improve the effectiveness of SafeSpeech by two measures. First, since the generation of perturbations is constrained by $\ell_p$ norm, increasing the perturbation radius can yield better effects, as proven by the experiment in Appendix \ref{section_append_radius}. Simultaneously, we can improve the acceptability of the perturbation by proposed STFT and SIOT metrics. Secondly, utilizing the surrogate model can be considered. From the experiment in Appendix \ref{section_append_surrogate}, we can find that the specific perturbations generated on this model slightly outperform transferability-based samples. Therefore, to improve the effectiveness of unknown models, an ensemble of models can be employed as~\cite{yu2023antifake}, although this will come with a significant computational cost.

\noindent\textbf{Broader Protective Strength.~} (1) {\it \uline{Effectiveness}}. We have evaluated the effectiveness of SafeSpeech under fine-tuning and zero-shot scenarios on single and multiple-speaker datasets. 
(2) {\it \uline{Transferability}}. We use one surrogate model to protect the dataset and validate the transferability across the other ten models. (3) {\it \uline{Robustness and Real-Time}}. We have considered a wide range of robustness in data, model, and real-world levels and confirm the real-time capability under speaker $\to$ microphone $\to$ SafeSpeech $\to$ speaker chain.

\noindent\textbf{Long and Complex Audio.} In daily utilization, users may aim to protect much longer audio samples. SafeSpeech has scalability and can handle longer and more complex audio. In Section \ref{section_robust_real-world}, the volunteer read audio that lasted about 10 minutes each test. Due to the universality of SafeSpeech, the generated perturbations can be scaled to longer audio. We have also demonstrated this point with the audio volume of approximately two hours for the two datasets in Section \ref{section_exp_finetuning}. Section \ref{section_robust_real-world} also demonstrates that the transmission time of the perturbations on different devices is only 0.369 seconds each time, which is acceptable.
For more complex audio, \textit{e.g.}, containing significant noise, TTS models tend to produce lower-quality outputs even without protection by SafeSpeech.

\section{Related Work}\label{section8}
\subsection{Audio Privacy Preservation}\label{section7.2}
Current voice privacy-enhancing techniques also include speaker anonymization~\cite{fang2019speaker, miao2023speaker} and audio watermarking~\cite{zhang2022robust}.
Speaker anonymization aims to protect the speaker's identity in voice data while preserving the speech content~\cite{srivastava2020evaluating}. Physical anonymization~\cite{hashimoto2016privacy} can be used to isolate the original speech, while logical anonymization~\cite{yao2024distinctive} bypass the authentication system.
Audio watermarking aims to protect the audio copyright~\cite{natgunanathan2022blockchain}, content authentication~\cite{alsabhany2024lightweight}, timbre certification~\cite{timbrewatermarking-ndss2024}, \textit{etc.}, without altering the original audio quality by embedding specific information. However, these methods cannot achieve SA2 with high-quality synthesis.

\subsection{Perturbative Availability Poisons}\label{section7.3}
PAP techniques are designed to prevent models from learning (\textit{e.g.}, by adding perturbations to the data)~\cite{huang2021unlearnable, liu2023image, PUE}.
Huang \textit{et al.}~\cite{huang2021unlearnable} found that the model learns the embedded error-minimizing noise rather than the information on clean labels. Building on this, Fowl \textit{et al.}~\cite{fowl2021adversarial} utilized adversarial examples for more effective data poisoning. Yu \textit{et al.}~\cite{yu2022availability} generated linearly separable Gaussian perturbations in the $\ell_ 2$ plane. These approaches have achieved SOTA PAP in the classification tasks~\cite{liu2023image}. Models may incorrectly assume that the data is not worth learning during the learning process due to the effect of specific clean-label noise, and thus discard noisy data due to fitting problems, resulting in unlearnable datasets. However, unlearnable examples~\cite{huang2021unlearnable} are fragile to data augmentation~\cite{ wu2023onepixel} and robust training~\cite{fu2022robust}, which can weaken the poisoning effect on unlearnable examples.

\section{Conclusion}\label{section9}
In this paper, we propose a proactive defense framework, SafeSpeech, to protect our voices from unauthorized speech synthesis via embedding imperceptible perturbations on original speeches before publicly releasing them. Extensive experimental evaluation shows that SafeSpeech has the most advanced voice protection effect to date, sufficient transferability to face various TTS models with distinct structures and backbones, and can resist the strength of various adaptive attackers in the real world. Moreover, SafeSpeech can effectively achieve real-time voice protection under scenarios of personal on-site presentations and reduce the security threats brought by voice cloning in the real world.

\section*{Acknowledgements}
We sincerely thank the anonymous reviewers and the shepherd for their constructive feedback on our work. This research is supported in part by the National Natural Science Foundation of China under Grant No. U21B2020, the Beijing Natural Science Foundation under Grant No. QY24213, the Fundamental Research Funds for the Central Universities under Grant No. 2024ZCJH05, and the National Natural Science Foundation of China under Grant No. 62202064.

\section*{Ethics Considerations}
We pay great attention to the potential safety issues that various research in society may raise, including those arising during the experimental phase and the release of SafeSpeech for use. In this paper, we strive to mitigate ethical concerns.

\noindent\textbf{Subjective Consideration.} A crucial purpose of our method is to prevent individuals from being deceived by deepfake audio, making human interaction experiments particularly important. Before conducting experiments, we considered ethical implications and sought opinions from relevant entities. The primary author's affiliated Human Ethics Research Committee determined that this study was exempt from further human subject review. All participants are over 18 years old, and we have sought their consent before experimenting. Throughout the experiment, no additional information was collected from the participants; all responses were anonymized. At the same time, we have informed them in advance that the content in these audios (LibriTTS dataset~\cite{zen2019libritts}) comes from an audiobook, and it is not a real event. Finally, the fake audio produced by this experiment (especially high-quality usable audio) is only used for the research of this experiment, not for other research, and these deepfake speeches were abandoned after the research was carried out.

\noindent\textbf{Legitimate and Beneficial Usage.} The initial intention of designing SafeSpeech is to prevent malicious speech synthesis and protect personal voice privacy. However, we recognize that speech synthesis can also be legitimate and beneficial, such as for disabled individuals who require speech synthesis tools. Therefore, SafeSpeech should not impede positive speech synthesis. Experimental results detailed in Section \ref{section_robustness_finetunig} demonstrate that training with a mix of protected and clean audio does not affect the synthesis quality of unprotected voices by SafeSpeech.
Moreover, we will release SafeSpeech by authorized request. If users want to utilize SafeSpeech to protect their voices, they need to obtain our written usage authentication and fill in the usage rules of SafeSpeech, which state that legitimate and beneficial uses of TTS tools are not allowed to be perturbed. Meanwhile, they should sign relevant \textbf{disclaimer clauses} to ensure that their usage behaviors are not related to the designer and publisher of SafeSpeech.

\section*{Open Science}
Before commencing the experiments, we are grateful for the open-source nature of the software and dataset used in this work and have taken into account the benefits that the principles of open science bring to research. Therefore, we have opened
our source code, datasets, and pre-trained models on \href{https://zenodo.org/records/14736906}{https://zenodo.org/records/14736906}, accompanied by a detailed description, \textit{e.g.}, the \texttt{README} file. The datasets and models we used are all open-source files, with no proprietary datasets or models, and we have provided references or links to pre-trained models for indexing.

% \clearpage
\normalem
\bibliographystyle{plain}
\bibliography{main}

\clearpage

\appendix

\begin{table*}[t]
    \centering
    \caption{Comparison of SafeSpeech and related works.}
    \setlength\tabcolsep{2pt}
    \setlength{\extrarowheight}{5pt}  
    \resizebox{0.98\textwidth}{!}{
    \begin{threeparttable}
    \begin{tabular}{c|c|c|c|c|c|c|c|c}
        \hline
        \textbf{Method} & \textbf{Type} & \textbf{E2E} & \textbf{Stage} & \textbf{Target Task} & \textbf{Transferability} & \textbf{Imperceptibility} & \textbf{Real-Time} & \textbf{Application Scenario} \\
        \hline
        \makecell[c]{Unlearnable \\ Examples}~\cite{huang2021unlearnable} 
            & \multirow{5}{*}{
                \makecell[c]{Perturbative \\ Availability \\ Poison}}
            & \multirow{4}{*}{\Checkmark}
            & \multirow{5}{*}{Training}
            & \multirow{5}{*}{
                \makecell[c]{image \\ classification}}
            & \multirow{3}{*}{\XSolidBrush}
            & \multirow{4}{*}{$\ell_{\infty}$ norm} 
            & \multirow{6}{*}{\XSolidBrush}
            & \multirow{5}{*}{
                \makecell[c]{Protect data \\ against authorized training.}}\\
        \cline{1-1}

        AdvPoison~\cite{fowl2021adversarial} & & & & & & & \\
        \cline{1-1}\cline{6-6}

        SEP~\cite{chen2023selfensemble} & & & & & checkpoint ensemble & & \\
        \cline{1-1}\cline{3-3}\cline{6-7}

        PTA~\cite{huang2021unlearnable,kim2021conditional} &
            & \multirow{2}{*}{\XSolidBrush} & &
            & \multirow{3}{*}{\XSolidBrush}
            & patch segment &   \\
        \cline{1-2}\cline{4-5}\cline{7-7} \cline{9-9}

        AttackVC~\cite{huang2021defending} 
            & \multirow{3}{*}{
                \makecell[c]{Voice Protection of \\ Identification }} &
            & \multirow{3}{*}{Inference} & voice conversion & 
            & \multirow{2}{*}{$\ell_{\infty}$ norm} &
            & \multirow{4}{*}{\makecell[c]{
                Protect personal \\voice information against \\malicious voice cloning.}}  \\
        \cline{1-1} \cline{3-3} \cline{5-5}\cline{8-8}

        VSMask~\cite{wang2023vsmask} &  
            & \multirow{2}{*}{\Checkmark} &
            & \multirow{2}{*}{\makecell[c]{zero-shot\\speech synthesis}} & & 
            & \Checkmark  \\
        \cline{1-1}\cline{6-8}

        AntiFake~\cite{yu2023antifake} & & & & & encoder ensemble
            & Frequency Penalty and SNR & \XSolidBrush & \\
        \cline{1-8}

        \textbf{\makecell[c]{SafeSpeech \\ (ours)}} 
            & \makecell[c]{
                Voice Protection of \\ Synthesis Quality and \\ Identification}
            & \Checkmark
            & Training 
            & \makecell[c]{zero-shot and \\ fine-tuning \\ speech synthesis}
            & \makecell[c]{pivotal and universal \\objective optimization}
            & \makecell[c]{STFT and STOI metrics \\ (time and frequency domain)} 
            & \Checkmark & \\

        \hline
    \end{tabular}
    \begin{tablenotes}
        \item (1){\textbf{Transferability}: The improvement techniques of transferability. (2) \textbf{E2E}: ``End-to-End'' represents whether the perturbation is embedded into the entire waveform not in the latent space. (3) The source code of VSMask is not public and unavailable.}
    \end{tablenotes}
    \end{threeparttable}
    }\label{tab_baseline}
\end{table*}

\section{Summary and Comparison of Related Work} \label{section_work_comparison}

In Table \ref{tab_baseline}, we describe the comparison between SafeSpeech and related works, including type, whether the perturbation is end-to-end, protection stage, target task, transferability enhancement technique, imperceptibility method, real-time capability, and application scenario. Previous protection techniques can mainly be classified into two categories, {\it \uline{perturbative availability poisons}} and {\it \uline{voice protection}}. PAP safeguards data by preventing its unauthorized use in model training. ``Availability'' in the PAP means that the source data cannot be used for training purposes~\cite{yu2022availability}. Voice protection employs adversarial examples to prevent voice cloning at the inference stage and protect personal identification.

\section{Details of Datasets and Models}\label{sectionA}
In this section, we introduce the principles of speaker selection and provide detailed information on datasets and models.

\begin{table}[H]
    \centering
    \caption{The detailed information of our selected datasets.}
    \resizebox{\linewidth}{!}{
    \setlength{\extrarowheight}{5pt}
    \begin{threeparttable}
    \begin{tabular}{c|c|c|c|c|c}
    \hline
    Name & Nums & Speakers (M/F) & Sampling Rate & 
    Max/Min Length & Average Length \\
    \hline
    $D_1$ &134 &1 (0/1) &24000  &9.94/0.61 (s) & 4.51 (s)  \\
    \hline
    $D_2$ &1800 &18 (11/7) &16000 & 6.69/1.20 (s) & 3.15 (s) \\
    \hline
    \end{tabular}
    \begin{tablenotes}
        \item ${D_1}$ and ${D_2}$ represent the dataset LibriTTS and CMU ARCTIC respectively.
    \end{tablenotes}
    \end{threeparttable}
    }
    \label{table_dataset_info}
\end{table}

\noindent{\textbf{Dataset Information}}. Some details about the two datasets we selected are shown in Table \ref{table_dataset_info} including numbers of samples, speaker categories, sampling rate of the dataset, maximum length, minimum length, and average length of dataset.

\begin{table}[H]
    \centering
    \caption{The detailed information of our selected models.}
    \setlength\tabcolsep{2pt}
    \resizebox{\linewidth}{!}{
    \setlength{\extrarowheight}{5pt}
    \begin{threeparttable}
    \begin{tabular}{c|c|c|c|c|c|c}
    \hline
    \textbf{Models} & \textbf{Type} & \textbf{Structure} & \textbf{Vocoder}
        & \textbf{LLM} & \textbf{NAR} & \textbf{Hours} \\
    \hline
    BERT-VITS2~\cite{github:Bert-VITS2} & fine-tuning & LLM + VAE & HiFiGAN~\cite{kong2020hifi}
        & DeBERTaV3~\cite{he2023debertav} & - & - \\
    \hline
    FishSpeech~\cite{liao2024fish} 
        & \multirow{2}{*}{zero-shot} 
        & \makecell[c]{LLM + Dual-AR} & Firefly-GAN
        & Llama~\cite{touvron2023llama1} 
       & \XSolidBrush & 720K \\
    \cline{1-1} \cline{3-7}
    F5-TTS~\cite{chen2024f5} 
        & & flow matching & Vocos~\cite{siuzdak2023vocos}
        & \multirow{7}{*}{-} 
        & \multirow{5}{*}{\Checkmark} & 95K \\
    \cline{1-4} \cline{7-7}
    GlowTTS~\cite{kim2020glow} 
        & \multirow{2}{*}{fine-tuning} & flow-based 
        & \multirow{3}{*}{HiFiGAN~\cite{kong2020hifi}}  & & & 24 \\
    \cline{1-1} \cline{3-3} \cline{7-7}
    MB-iSTFT-V~\cite{kawamura2023lightweight} 
        & & iSTFT + VAE & & &  & 24 \\
    \cline{1-3} \cline{7-7}
    OpenVoice~\cite{qin2023openvoice} 
        & zero-shot & encoder-decoder &&&& 3.5K \\
    \cline{1-4} \cline{7-7}
    StyleTTS 2~\cite{li2024styletts} 
        & fine-tuning & diffusion-based 
        &\makecell[c]{HiFiGAN~\cite{kong2020hifi} \\ iSTFTNet~\cite{kaneko2022istftnet}} &&& 245\\
    \cline{1-4} \cline{6-7}
    TorToise-TTS~\cite{betker2023better} 
        & zero-shot & diffusion-based & UnivNet~\cite{univnet} & & \XSolidBrush & 46K \\
    \cline{1-4} \cline{6-7}
    VITS~\cite{kim2021conditional} 
        & fine-tuning & VAE & \multirow{2}{*}{HiFiGAN~\cite{kong2020hifi}} & & \Checkmark & 24 \\
    \cline{1-3} \cline{5-7}
    XTTS~\cite{casanova2024xtts} 
        & zero-shot & LLM + VQ-VAE &
        & GPT-2~\cite{radford2019language}
        & \XSolidBrush & 27K \\
    \hline
    \end{tabular}
    \begin{tablenotes}
        \item (1){\textbf{VQ-VAE}: Vector Quantized-Variational AutoEncoder. 
            (2) \textbf{Dual-AR}: Dual Autoregressive. 
            (3) \textbf{iSTFT}: inverse Short-Time Fourier Transform.
            (4) \textbf{FireflyGAN}: FireflyGAN is an enhanced vocoder proposed in the FishSpeech.}
    \end{tablenotes}
    \end{threeparttable}
    }
    \label{table_models}
\end{table}

\noindent\textbf{Model Information.}
To facilitate a more comprehensive comparison of the various models, we show the Table \ref{table_models}. This table highlights differences among models across several key dimensions: model type, structure, incorporation of the LLM component, non-autoregressive (NAR) status, and the volume of training dataset for the base models. In cases where models do not incorporate the LLM component, this is indicated with a dash (``-'') in the corresponding ``LLM'' column. It should be noted that for BERT-VITS2, official disclosure regarding the volume of training data is unavailable.

\section {Details of User Study}\label{section_append_user}
\noindent\textbf{Filtering.} To ensure participant seriousness, we designed two simple English arithmetic questions in the questionnaire at random positions. Incorrect answers to these questions indicated a lack of seriousness, leading to the exclusion of those responses. Additionally, we filtered out participants who provided identical or random answers throughout.

\noindent\textbf{Experimental Generalizability.} To minimize the bias introduced by subjective experiments and enhance the generalizability of the survey, we have employed several techniques. (1) {\it \uline{Adequate Participants}}. Compared to related works such as AntiFake~\cite{yu2023antifake}, which used 24 participants,  and VSMask~\cite{wang2023vsmask}, which used 25 participants, we invited 80 individuals to take part in our survey. (2) {\it \uline{Randomization and Anonymization}}: The order of questions in the questionnaire was completely randomized. We provided no hints or additional instructions within the questions, and participants had no way of knowing which algorithm generated the audio they were currently listening to. (3) {\it \uline{Confidence Interval}}: Considering the potential bias, \textit{e.g.}, personal preferences of participants, in result calculations of the subjective survey, we computed a 95\% confidence interval for MOS and naturalness values to provide more reliable results as Eq. (\ref{eq_mos_1}) and (\ref{eq_mos_2}).

We assume that the MOS score of model $i$ is $\mu_i$, and in addition, the 95\% confidence interval~\cite{kumar2019melgan} score is $CI_i$, which can be calculated using the following formula:
\begin{equation}
    \hat{\mu_i}=\frac{1}{N_i}\cdot\sum\limits_{k=1}^{N_i}m_{i,k},\label{eq_mos_1}
\end{equation}
\begin{equation}
    CI_i=\left[\hat{\mu_i}-1.96\frac{\hat{\sigma_i}}{\sqrt{N_i}}, \hat{\mu_i}+1.96\frac{\hat{\sigma_i}}{\sqrt{N_i}} \right],\label{eq_mos_2}
\end{equation}
where $\hat{\sigma_i}$ is the standard deviation of the scores collected.

\noindent\textbf{Rating Principles} In the questionnaire, participants listened to each audio and rated its quality from 0 to 5 based on their subjective perception. A score of 5 indicates excellent audio quality (smooth and noise-free). 4 suggests good quality (minimal noise and delays, easy to understand). 3 means average quality (with noise and delays but understandable). 2 denotes fairly poor quality (requiring multiple repetitions to understand). 1 indicates poor quality (very hard to understand), and 0 represents extremely poor quality (completely inaudible). 

\begin{table}[t]
    \centering
    \caption{The transferability performance when regarding MB-iSTFT-VITS as our surrogate model.}
    \resizebox{0.9\linewidth}{!}{
    \begin{tabular}{cccccccc}
        \toprule
        \multirow{2}{*}{Method}
        & \multicolumn{3}{c}{BERT-VITS2} & \multicolumn{3}{c}{MB-iSTFT-VITS} \\
        \cmidrule(r){2-4}\cmidrule(lr){5-7}
        & MCD($\downarrow$)   & WER(\%)($\downarrow$)  & SIM($\uparrow$)
        & MCD($\downarrow$)   & WER(\%)($\downarrow$)& SIM($\uparrow$)\\
        \midrule
        clean &5.099 &25.095 &0.625 &5.139 &20.913 &0.623 \\
        PTA &9.949 &59.646 &0.266 &8.892 &47.569 &0.219 \\
        \textbf{SPEC (ours)}&\textbf{12.791} &\textbf{93.552} &\textbf{0.215} &\textbf{12.374} & \textbf{124.142} & \textbf{0.159}\\
        \bottomrule
    \end{tabular}
    }
\label{tab_surrogate}
\end{table}
  
\section{Additional Experiments}\label{sectionD} 
In this section, we conduct an easier fine-tuning model, leveraging a web interface operation. Additionally, we study the impact of the surrogate model and noise radius of the $\ell_p$ norm.

\subsection{WebUI Operation} \label{sectionD.1}

Recently, an efficiently fine-tuning TTS synthesis model named GPT-SoVITS~\cite{github:GPT-SoVITS} has garnered over \textbf{38K} stars in the GitHub community, which supports WebUI-based training.

GPT-SoVITS has received numerous positive feedback as public users are amazed at its convenience, high performance, and efficient training operation. It consists of GPT~\cite{radford2019language} and a VITS-based synthesizer, utilizing an LLM component to make the synthesizer understand the text better. The two parts are trained separately.
To reproduce the attackers' training process, we use the web-based training method provided by the authors. We upload SafeSpeech-protected audio to the designated path and do audio size division, automatic text recognition, and annotation.  Unlike previous experiments where training text was manually annotated, we employ Whisper in Section \ref{section_settings_metrics} to recognize text. Then we set the training iterations for SoVITS and GPT as 25 and 50 respectively and evaluate the performance on the web-based page.

The result shows that the SIM metric of the synthetic speeches is only 0.25, meaning the dissimilarity between synthetic and original speeches. In this experiment, we fine-tune a well-known and advanced TTS model by web-based operation, which ensures that we have not modified the model and simulate a possible training scenario in the real world.

\subsection{Alternative Surrogate Model Choice}\label{section_append_surrogate}

SafeSpeech protects datasets based on the surrogate model. In previous experiments, we choose BERT-VITS2 as the surrogate model in our previous experiments, while our methods have no limitation on model selection. Therefore, the user can utilize the specific model for specific scenarios, \textit{e.g.}, MB-iSTFT-VITS in real-time applications.

Table \ref{tab_surrogate} shows the experimental results when regarding MB-iSTFT-VITS as the surrogate model. We can find that after training on the protected dataset, the value WER of 124.142\% means highly unclear synthesized audio, with a speaker similarity from 0.623 trained on the clean samples to 0.159, and the ASR of the speech synthesis attack is only 3.846\% representing a successful defense. Moreover, the perturbation is also transferable on BERT-VITS2. This experiment serves as an excellent demonstration of the versatility of SafeSpeech in design, that noise generation does not depend on a specific model with high effectiveness and transferability.

\begin{table}[t]
    \centering
    \caption{Quantitative analysis on different perturbation boundary $\epsilon$ with $4/255$ and $16/255$ on BERT-VITS2 model.}
    \resizebox{0.9\linewidth}{!}{
    \begin{tabular}{cccccc ccccc}
      \toprule
      \multirow{2}[1]{*}{Methods}
      & \multicolumn{3}{c}{$4/255$} 
      & \multicolumn{3}{c}{$16/255$} 
      \\
      \cmidrule(r){2-4}\cmidrule(l){5-7}
      & MCD($\uparrow$)   & WER(\%)($\uparrow$)  & SIM($\downarrow$)   
      & MCD($\uparrow$)   & WER(\%)($\uparrow$)  & SIM($\downarrow$)\\
      \midrule
        clean  
            &5.099 &25.095 &0.625 &5.099 &25.095 &0.625\\
        random noise 
            &5.334 &34.575 &0.584 &6.614 &43.142 &0.433\\
        AdvPoison~\cite{fowl2021adversarial} 
            &7.059 &46.030 &0.403 
            &13.718 &\uline{91.525} &\uline{0.205} \\
        SEP~\cite{chen2023selfensemble} 
            &6.084 &44.926 &0.401 &12.737 &83.804 &0.263 \\
        PTA~\cite{huang2021unlearnable} 
            &\uline{7.360} &\uline{49.553} &\uline{0.342}
            &\uline{17.040} &89.875 &0.206 \\
        \textbf{SPEC (ours)} 
            &\textbf{10.921} &\textbf{76.110} &\textbf{0.302} 
            &\textbf{18.634} &\textbf{105.385} &\textbf{0.093} \\
      \bottomrule
    \end{tabular}
    }
    \label{tab_epsilon}
\end{table}

\subsection{Perturbation Boundaries}\label{section_append_radius}

The perturbation radius $\epsilon$ plays an important role in the $\ell_p$ norm constraints of SafeSpeech. Higher $\epsilon$ can achieve better effects but worse perception. Table \ref{tab_epsilon} presents the results when $\epsilon$ is $4/255$ and $16/255$, respectively. When $\epsilon$ takes $16/255$, it can be found that the data protection effect is excellent, with SIM of only 0.093 and attack success rate of 0\%, and WER of 105.385\%, which means a successful defense. If users aim to realize stronger strength to mitigate the circumvention of SafeSpeech, larger $\epsilon$ can be set, containing the perceptual optimization in Section \ref{section_method_perception}.

\end{document}